\documentclass[pdflatex,sn-nature]{sn-jnl}

\usepackage{graphicx}%
\usepackage{multirow}%
\usepackage{amsmath,amssymb,amsfonts}%
\usepackage{amsthm}%
\usepackage[title]{appendix}%
\usepackage{xcolor}%
\usepackage{textcomp}%
\usepackage{manyfoot}%
\usepackage{booktabs}%
\usepackage{algorithm}%
\usepackage{algorithmicx}%
\usepackage{algpseudocode}%
\usepackage{listings}%
\usepackage{upgreek}
\usepackage{xspace}
\usepackage{caption}

\newcommand{\um}{$\upmu$m\xspace}

\begin{document}
\newcounter{firstbib}
\title{An inspiral of a star inside a common envelope with its giant companion}

\let\oldabstract\abstract
\renewcommand{\abstract}[1]{\oldabstract{\mathversion{normal}#1}}

\author*[1,2]{\fnm{Viraj} \sur{Karambelkar}}
\email{vk2588@columbia.edu}
\author[1,3]{\fnm{Kishalay} \sur{De}}
\author[4]{\fnm{Karel} \spfx{D.} \sur{Temmink}}
\author[3,5]{\fnm{Brian} \spfx{D.} \sur{Metzger}}
\author[6,7,8]{\fnm{Nadejda} \sur{Blagorodnova}}
\author[9]{\fnm{G.} \spfx{C.} \sur{Anupama}}
\author[9]{\fnm{Sudhanshu} \sur{Barway}}
\author[9]{\fnm{Judhajeet} \sur{Basu}}
\author[11]{\fnm{Eric} \spfx{C.} \sur{Bellm}}
\author[12]{\fnm{Varun} \sur{Bhalerao}}
\author[13]{\fnm{Aleksandra} \sur{Bochenek}}
\author[9,10]{\fnm{Shatakshi} \sur{Chamoli}}
\author[14]{\fnm{Kaustav} \sur{Das}}
\author[15]{\fnm{Christoffer} \sur{Fremling}}
\author[16]{\fnm{Danielle} \sur{Frostig}}
\author[6,7,8]{\fnm{Marco} \spfx{A.} \sur{Gómez-Muñoz}}
\author[17]{\fnm{Steven} \spfx{L.} \sur{Groom}}
\author[14]{\fnm{K-Ryan} \sur{Hinds}}
\author[14]{\fnm{Joahan} \spfx{Castaneda} \sur{Jaimes}}
\author[17]{\fnm{Jacob} \sur{Jencson}}
\author[18,19]{\fnm{David} \sur{Jones}}
\author[14]{\fnm{Mansi} \sur{Kasliwal}}
\author[6,7]{\fnm{Grace} \sur{Katusiime}}
\author[31]{\fnm{Rubina} \sur{Kotak}}
\author[14]{\fnm{S. R.} \sur{Kulkarni}}
\author[17]{\fnm{Ryan} \sur{Lau}}
\author[20,21,22]{\fnm{Chang} \sur{Liu}}
\author[23]{\fnm{Nathan} \spfx{P.} \sur{Lourie}}
\author[16]{\fnm{Morgan} \sur{MacLeod}}
\author[32]{\fnm{Keiichi} \sur{Maeda}}
\author[13]{\fnm{Zoe} \sur{McGrath}}
\author[14,24]{\fnm{Geoffrey} \sur{Mo}}
\author[6,7,8]{\fnm{Gerard} \sur{Garcia-Moreno}}
\author[15]{\fnm{Josiah} \sur{Purdum}}
\author[20,21,22]{\fnm{Nabeel} \sur{Rehemtulla}}
\author[25]{\fnm{R. Michael} \sur{Rich}}
\author[15]{\fnm{Reed} \spfx{L.} \sur{Riddle}}
\author[14]{\fnm{Sam} \sur{Rose}}
\author[23]{\fnm{Robert} \spfx{A.} \sur{Simcoe}}
\author[26]{\fnm{Jesper} \sur{Sollerman}}
\author[27,28,29]{\fnm{Robert} \sur{Stein}}
\author[6,7,8]{\fnm{Hugo} \sur{Tranin}}
\author[6,7,8]{\fnm{Maxime} \sur{Wavasseur}}
\author[30]{\fnm{Simon} \spfx{de} \sur{Wet}}
\author[13]{\fnm{Jacob} \sur{Wise}}

\affil[1]{\orgdiv{Columbia Astrophysics Laboratory}, \orgname{Columbia University}, \orgaddress{\street{538 West 120th Street 704, MC 5255}, \city{New York}, \state{NY}, \postcode{10027}, \country{USA}}}

\affil[2]{NASA Hubble Fellow}

\affil[3]{\orgname{Center for Computational Astrophysics, Flatiron Research Institute}, \orgaddress{\street{162 5th Ave}, \city{New York}, \state{NY}, \postcode{10010}, \country{USA}}}

\affil[4]{\orgdiv{Department of Astrophysics/IMAPP}, \orgname{Radboud University Nijmegen}, \orgaddress{\street{P.O. Box 9010}, \city{Nijmegen}, \postcode{6500 GL}, \country{The Netherlands}}}

\affil[5]{\orgdiv{Department of Physics and Columbia Astrophysics Laboratory}, \orgname{Columbia University}, \orgaddress{\city{New York}, \state{NY}, \postcode{10027}, \country{USA}}}

\affil[6]{\orgdiv{Institut de Ciències del Cosmos (ICCUB)}, \orgname{Universitat de Barcelona}, \orgaddress{\street{c. Martí i Franquès, 1}, \city{Barcelona}, \postcode{08028}, \country{Spain}}}

\affil[7]{\orgdiv{Departament de Física Quàntica i Astrofísica (FQA)}, \orgname{Universitat de Barcelona}, \orgaddress{\street{c. Martí i Franquès, 1}, \city{Barcelona}, \postcode{08028}, \country{Spain}}}

\affil[8]{\orgname{Institut d'Estudis Espacials de Catalunya (IEEC)}, \orgaddress{\street{Edifici RDIT, Campus UPC}, \city{Castelldefels}, \postcode{08860}, \country{Spain}}}

\affil[9]{\orgname{Indian Institute of Astrophysics}, \orgaddress{\street{II Block, Koramangala}, \city{Bengaluru}, \postcode{560 034}, \country{India}}}

\affil[10]{\orgname{Pondicherry University}, \orgaddress{\street{R.V. Nagar, Kalapet}, \city{Puducherry}, \postcode{605014}, \country{India}}}

\affil[11]{\orgname{DIRAC Institute}, \orgdiv{Department of Astronomy}, \orgname{University of Washington}, \orgaddress{\street{3910 15th Avenue NE}, \city{Seattle}, \state{WA}, \postcode{98195}, \country{USA}}}

\affil[12]{\orgdiv{Department of Physics}, \orgname{Indian Institute of Technology Bombay}, \orgaddress{\street{Powai}, \city{Mumbai}, \postcode{400076}, \country{India}}}

\affil[13]{\orgdiv{Astrophysics Research Institute}, \orgname{Liverpool John Moores University}, \orgaddress{\street{146 Brownlow Hill}, \city{Liverpool}, \postcode{L3 5RF}, \country{UK}}}

\affil[14]{\orgdiv{Division of Physics, Mathematics and Astronomy}, \orgname{California Institute of Technology}, \orgaddress{\street{1200 E. California Blvd}, \city{Pasadena}, \state{CA}, \postcode{91125}, \country{USA}}}

\affil[15]{\orgname{Caltech Optical Observatories, California Institute of Technology}, \orgaddress{\city{Pasadena}, \state{CA}, \postcode{91125}, \country{USA}}}

\affil[16]{\orgname{Center for Astrophysics | Harvard \& Smithsonian}, \orgaddress{\street{60 Garden Street}, \city{Cambridge}, \state{MA}, \postcode{02138}, \country{USA}}}

\affil[17]{\orgname{IPAC, California Institute of Technology}, \orgaddress{\street{1200 E. California Blvd}, \city{Pasadena}, \state{CA}, \postcode{91125}, \country{USA}}}

\affil[18]{\orgname{Instituto de Astrofísica de Canarias}, \orgaddress{\city{La Laguna}, \postcode{E-38205}, \country{Spain}}}

\affil[19]{\orgdiv{Departamento de Astrofísica}, \orgname{Universidad de La Laguna}, \orgaddress{\city{La Laguna}, \postcode{E-38206}, \country{Spain}}}

\affil[20]{\orgdiv{Department of Physics and Astronomy}, \orgname{Northwestern University}, \orgaddress{\street{2145 Sheridan Rd}, \city{Evanston}, \state{IL}, \postcode{60208}, \country{USA}}}

\affil[21]{\orgname{Center for Interdisciplinary Exploration and Research in Astrophysics (CIERA), Northwestern University}, \orgaddress{\street{1800 Sherman Ave}, \city{Evanston}, \state{IL}, \postcode{60201}, \country{USA}}}

\affil[22]{\orgname{NSF-Simons AI Institute for the Sky (SkAI)}, \orgaddress{\street{172 E. Chestnut St.}, \city{Chicago}, \state{IL}, \postcode{60611}, \country{USA}}}

\affil[23]{\orgdiv{MIT-Kavli Institute for Astrophysics and Space Research}, \orgname{Massachusetts Institute of Technology}, \orgaddress{\street{77 Massachusetts Ave.}, \city{Cambridge}, \state{MA}, \postcode{02139}, \country{USA}}}

\affil[24]{\orgname{The Observatories of the Carnegie Institution for Science}, \orgaddress{\city{Pasadena}, \state{CA}, \postcode{91101}, \country{USA}}}

\affil[25]{\orgdiv{Department of Physics and Astronomy}, \orgname{University of California, Los Angeles}, \orgaddress{\street{430 Portola Plaza, Box 951547}, \city{Los Angeles}, \state{CA}, \postcode{90095-1547}, \country{USA}}}

\affil[26]{\orgname{The Oskar Klein Centre}, \orgdiv{Department of Astronomy}, \orgname{Stockholm University}, \orgaddress{\street{AlbaNova}, \city{Stockholm}, \postcode{SE-10691}, \country{Sweden}}}

\affil[27]{\orgdiv{Department of Astronomy}, \orgname{University of Maryland}, \orgaddress{\city{College Park}, \state{MD}, \postcode{20742}, \country{USA}}}

\affil[28]{\orgname{Joint Space-Science Institute}, \orgaddress{\city{College Park}, \state{MD}, \postcode{20742}, \country{USA}}}

\affil[29]{\orgdiv{Astrophysics Science Division}, \orgname{NASA Goddard Space Flight Center}, \orgaddress{\street{Mail Code 661}, \city{Greenbelt}, \state{MD}, \postcode{20771}, \country{USA}}}

\affil[30]{\orgname{DTU Space, Technical University of Denmark}, \orgaddress{\street{Elektrovej 327--328}, \city{Lyngby}, \postcode{DK-2800}, \country{Denmark}}}

\affil[31]{\orgdiv{Department of Physics and Astronomy}, \orgname{University of Turku}, \orgaddress{\street{FI-20014}, \country{Finland}}}

\affil[32]{\orgdiv{Department of Astronomy}, \orgname{Kyoto University}, \orgaddress{\street{Kitashirakawa-Oiwake-cho, Sakyo-ku}, \city{Kyoto},\postcode{606-8502} \country{Japan}}}

\abstract{The formation of compact stellar binaries requires rapid orbital tightening of their progenitor binary stars \cite{Tauris2006,Postnov2014}. This is theorized to happen via Common Envelope Evolution (CEE, \cite{Paczynski1976IAUS, Webbink84, Podsiadlowski2003, Ivanova2013araa, Wilson2019}), when the stars orbit in a loosely bound, shared convective envelope, triggering an inspiral due to frictional drag \cite{Ivanova13}. However, this inspiral in such a system has never been observed directly. Here, we present observations of AT\,2025abao, a transient eruption in the Andromeda Galaxy, that originates from the CEE of a $\approx7$\,M$_{\odot}$ asymptotic giant branch (AGB) star with its companion. Extensive pre- and post-eruption spectroscopy is best explained as a $\approx0.7$\,M$_{\odot}$ companion inspiraling within the convective AGB envelope that powers a seven year period of gradual brightening -- the first direct observations of frictional drag energy dissipation induced by CEE. The subsequent merger of the companion with the primary core within a few ($<10$) orbits powers the ejection of the outermost $\approx10\%$ of the stellar envelope at speeds $\approx10\times$ the primary escape speed, and empirically constrains the fraction of dissipated orbital energy available to unbind the convective common envelope \cite{vandenHeuvel1976,Tutukov1979,Webbink84, Ivanova2013araa}, $0.01\lesssim \alpha \lesssim 0.4$. This end-to-end characterization using multi-survey observations demonstrates a path towards mapping the landscape of CEE outcomes with upcoming surveys.
}

\maketitle

\begin{figure}[H]
    \centering
    \includegraphics[width=\textwidth]{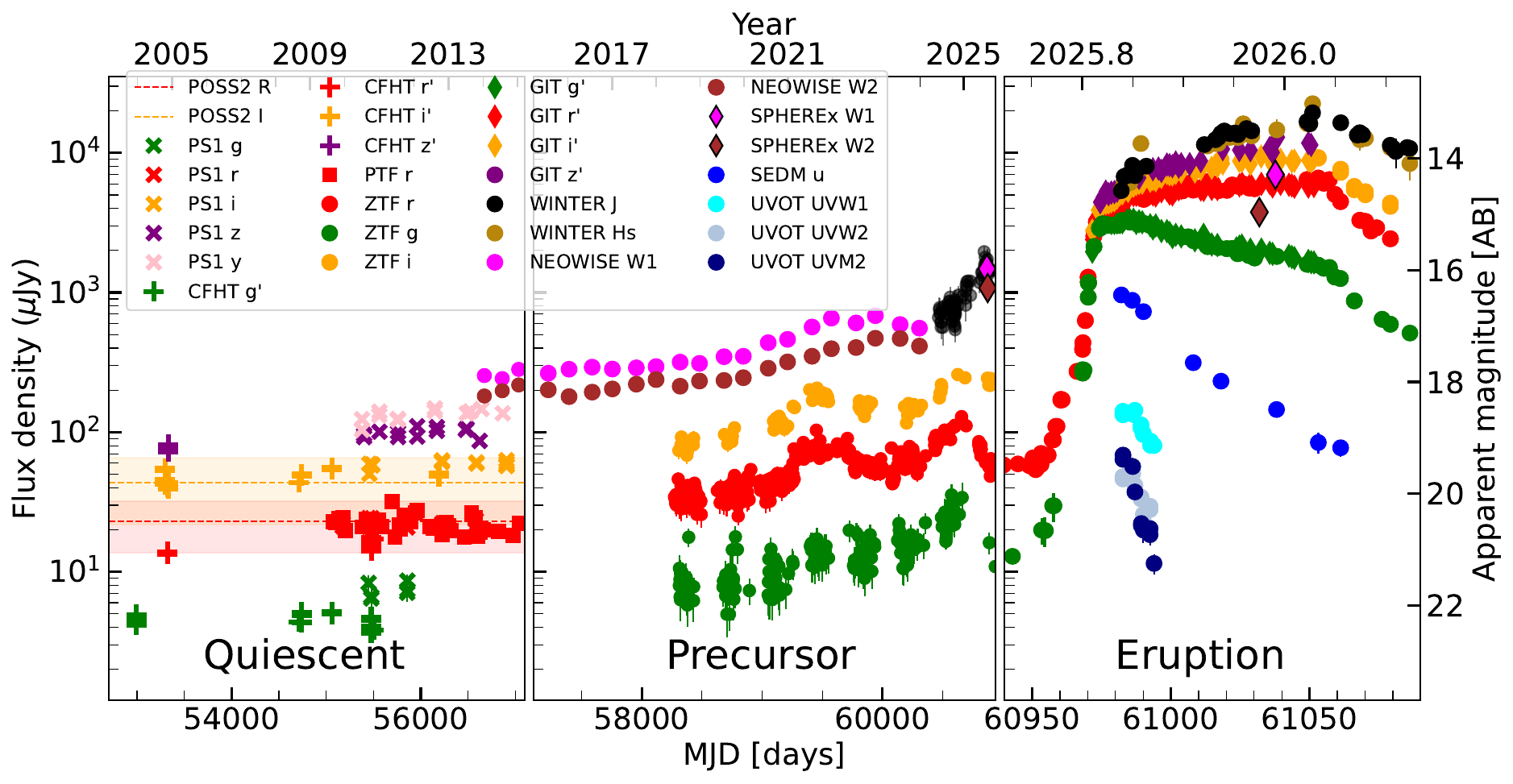}
    \caption{\textbf{Multiband lightcurve of AT\,2025abao} \\
    \small{Ultraviolet to mid-infrared lightcurve, constructed using POSS-2, PS1, CFHT, PTF, ZTF, GIT, WINTER, NEOWISE, SPHEREx, P60, and Swift data, with symbols and colors denoting different telescopes and filters (see legend and Section \ref{sec:followup_obs}). The three panels are continuous in time, but use different x-axis scales to illustrate the quiescent progenitor phase (2005 to 2015, left panel), the precursor phase (2015 to early 2025, middle panel), and the eruption (late 2025 to early 2026, right panel). The left panel also indicates fluxes from 1989 POSS R and I-band images (dotted red and orange lines, 1$\sigma$ uncertainties as shaded regions). SPHEREx spectrophotometry was converted to W1 and W2-band photometry (brown, magenta diamonds) by synthetic photometry over the spectrum using the respective transmission profiles.}}
    \label{fig:full_lightcurve}
\end{figure}

In April 2025, \cite{Karambelkar2025} reported the unusual and gradual brightening of an AGB star (named WNTR23bzdiq/WTP19aalzlk) in the Andromeda Galaxy. Multi-wavelength observations of this source indicated that this brightening may be powered by energy deposition of a companion spiraling through the star's outer convective envelope \cite{Karambelkar2025} -- marking the onset of common-envelope evolution (CEE \cite{Paczynski1976IAUS, Ivanova2013araa}), or possibly a brief brightening caused by nuclear energy input from thermal pulses in late-stage stellar evolution  \cite{Karambelkar2025}. The quiescent progenitor star shows no signs of variability between 1989 and the start of its brightening in 2018 (Figure \ref{fig:full_lightcurve}). Modeling the spectral energy distribution (SED) of the quiescent progenitor, \cite{Karambelkar2025} find that the source is consistent with a $\approx$7\,M$_{\odot}$ star in its early-AGB phase -- exhibiting a photospheric temperature of T$_{\rm{eff}}\approx3650$\,K, radius of R$_{\rm{eff}}\approx350$\,R$_{\odot}$, and total luminosity L $\approx1.8\times10^4$\,L$_{\odot}$ (Section \ref{sec:modeling_progenitor}). The large infrared excess in the SED indicates a wind mass loss rate of $\dot{M}\approx10^{-5}$\,M$_{\odot}$\,yr$^{-1}$ -- substantially higher than the expectations for isolated early AGB stars \cite{Reimers1975}, and supports a state of prolonged enhanced mass loss.

From 2018 to early 2025, the star brightened gradually, exhibiting two distinct episodes where its optical (0.4-0.8\um) flux increased by a factor of $2-3$ over a period of $\approx 1000-1500$\,days. The source subsequently faded rapidly by a factor of $\approx 2$ in optical flux for $\approx300$ days, even as its infrared (1-5\um) emission continued to brighten by a factor $\gtrsim 5$. Modeling the multi-color evolution, we find that the brightening phases sustained a near constant photospheric temperature of $\approx3800$\,K, before cooling to $\approx3500$\,K during the rapid optical fading (Section \ref{sec:modeling_precursor_sed}).

In July 2025, the source was serendipitously observed during its rapid optical fading by the first sky survey of the SPHEREx mission \cite{Bock2026}, providing a precise glimpse of the stellar photosphere during this phase. Modeling the SPHEREx SED (Section \ref{sec:modeling_spherex}) using models of dusty stellar atmospheres, we find that the stellar radius and luminosity have increased to R $\approx860$\,R$_{\odot}$ to L $\approx10^{5}$\,L$_{\odot}$ respectively, while the corresponding photospheric temperature has dropped to T $\approx3460$\,K. The SPHEREx data indicate a substantial increase in the circumstellar dust mass and mass loss rate relative to the progenitor star, corresponding to the ejection of $\sim0.1$\,M$_{\odot}$ of gas (for typical dust to gas ratios) and an increased mass loss rate of $\approx10^{-2}$\,M$_{\odot}$\,yr$^{-1}$ at this epoch.

\begin{figure}[H]
    \centering
    \includegraphics[width=\textwidth]{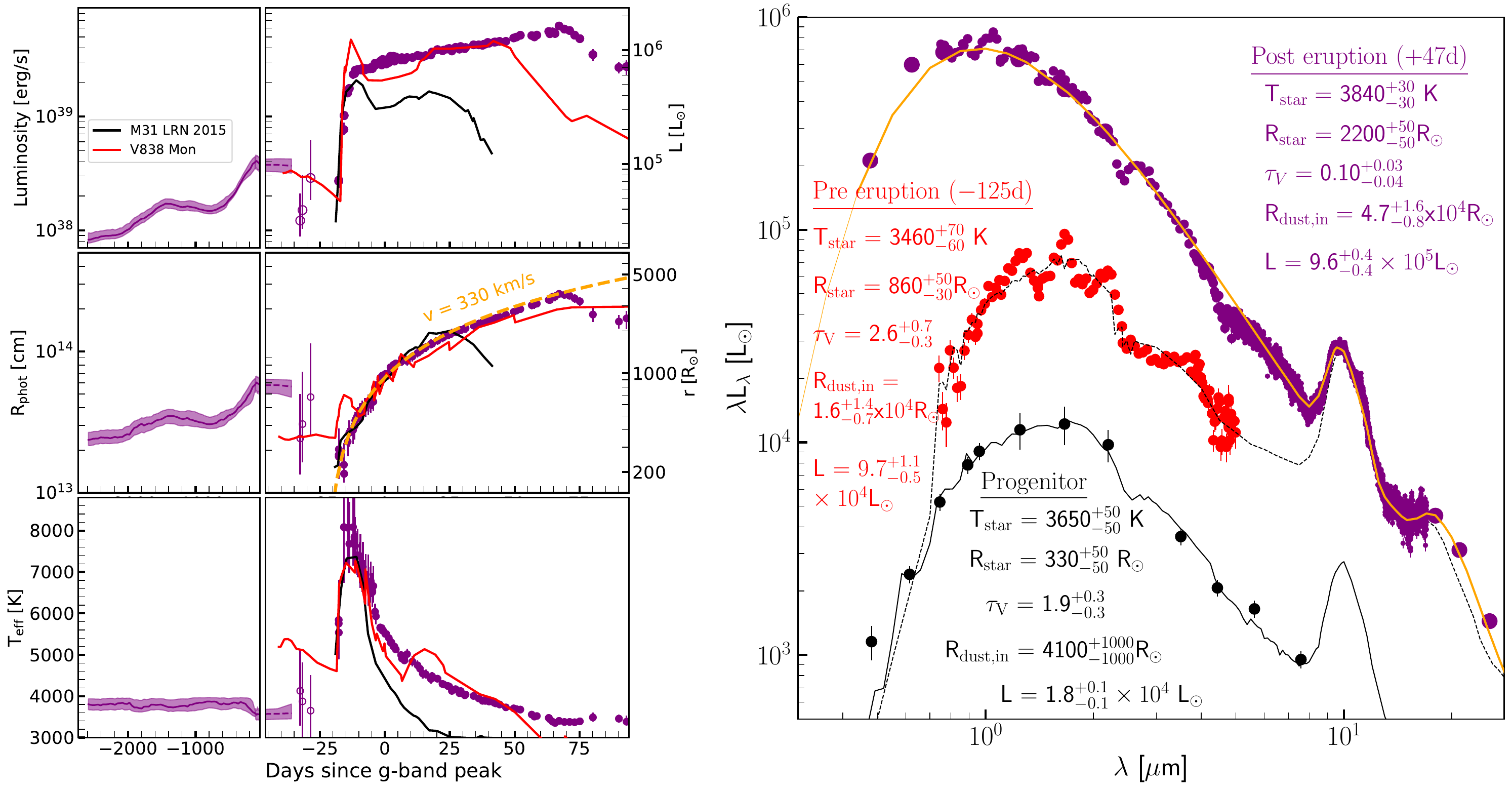}
    \caption{\textbf{Evolution of the bolometric and spectroscopic properties of AT\,2025abao}\\
    \small{\emph{Left:} Evolution of photospheric temperature (bottom), photospheric radius (middle), and bolometric luminosity (top). Purple lines and shaded regions (left panel) mark values and 1$\sigma$ uncertainties from fitting stellar models to precursor SEDs; purple dots (right panel) show those from blackbody fits to eruption SEDs. The two panels are contiguous in time but use different x-axis scales. Corresponding values of V838Mon (red) \cite{Munari2002AA} and M31-LRN-2015 (black) \cite{Williams2015, Kurtenkov15} are shown for comparison. The yellow dashed line (radius panel) shows a photosphere expanding at 330\,km\,s$^{-1}$. \\
    \emph{Right:} Quiescent progenitor SED (black points), precursor SPHEREx spectrum (red), and post-eruption SED from SPHEREx, \emph{JWST}, and ground-based data (purple), labeled by phase relative to the \emph{g-}band peak. Best-fit models are plotted as lines, and relevant inferred properties with 1$\sigma$ uncertainties are indicated.}}
    \label{fig:evolution_properties}
\end{figure}

\begin{figure}[H]
    \centering
    \includegraphics[width=\textwidth]{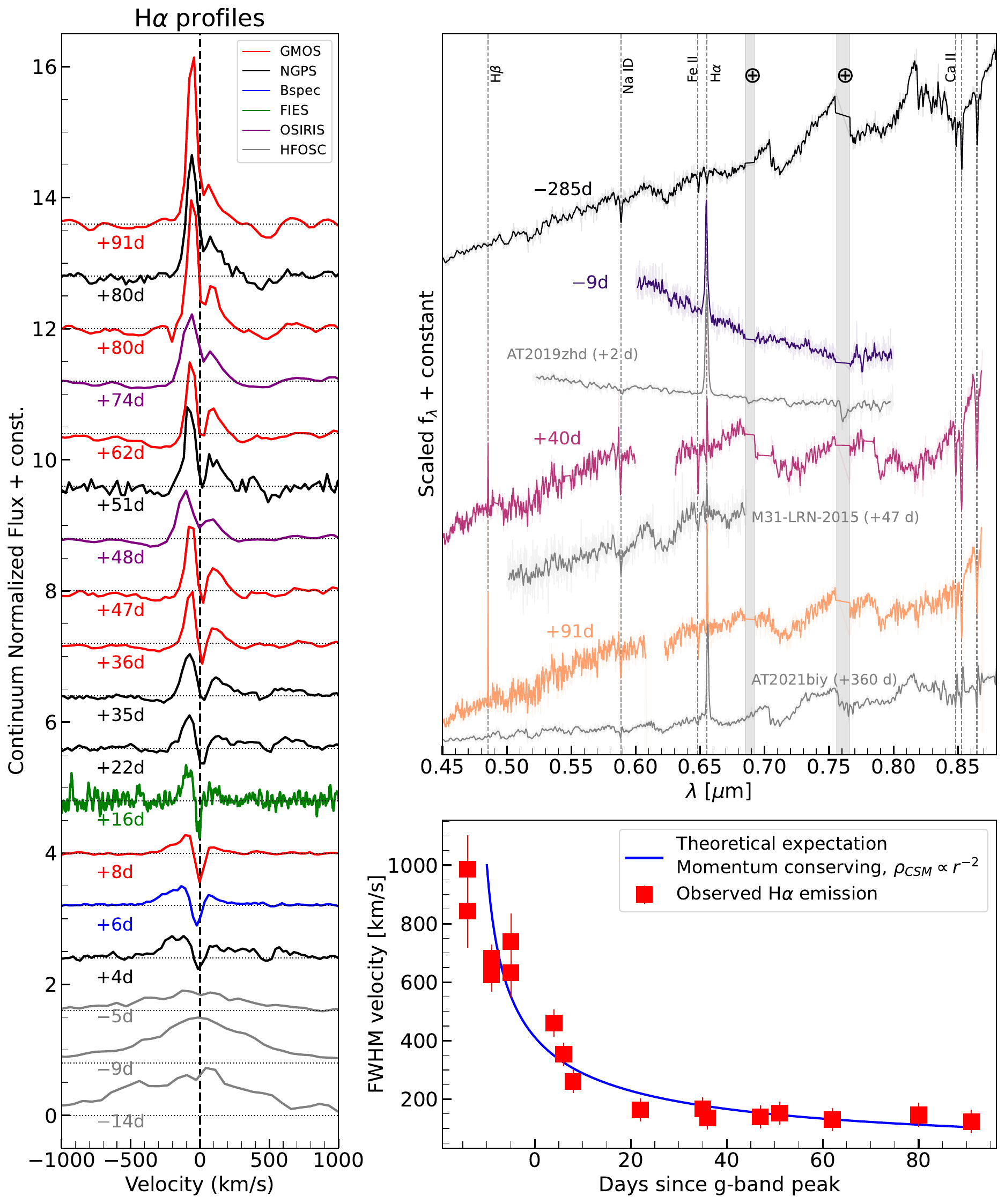}
    \caption{\textbf{Spectroscopic evolution and hydrogen line velocities of AT\,2025abao}\\
    \small{\emph{Left:} Continuum normalized H$\alpha$ spectral profiles colored by spectrograph and labeled by phase relative to the \emph{g-}band peak. Each spectrum is offset by a constant (dotted horizontal lines). The dashed vertical line marks zero expansion velocity. 
    \emph{Top right:} Representative optical spectra of the AT\,2025abao eruption from 285 days before to 91 days after the \emph{g-}band peak, compared with LRNe AT2019zhd \cite{Karambelkar2023}, M31-LRN-2015 \cite{Blagorodnova2020}, and AT2021biy \cite{Karambelkar2023} at similar phases. 
    \emph{Bottom right: } Evolution of FWHM velocities (and 1$\sigma$ uncertainties) of the H$\alpha$ emission inferred using gaussian fits, compared with the theoretical velocity deceleration expected from momentum conserving interactions between fast-moving ejecta and CBM (\cite{Metzger2017}, blue).}}
    \label{fig:opt_spec}
\end{figure}

In October 2025, the star underwent a dramatic eruption \cite{Taguchi2025ATel, Reguitti2026} immediately following the rapid optical fading -- brightening by a factor of $\approx100$ in $r-$band over 25 days and exhibiting a sudden increase in its photospheric temperature to $\approx 8000$\,K. Over the next eighty days, the luminosity plateaued, while the effective temperature steadily cooled to $\approx3000$\,K (Figure\,\ref{fig:evolution_properties}, Section\,\ref{sec:modeling_lrn_phot}). The peak bolometric luminosity during the eruption reached $\approx1.5\times10^{6}$L$_{\odot}$, and was accompanied by a steady photospheric expansion speed of $\approx330$\,km\,s$^{-1}$. Spectra obtained at early times during the eruption \cite{Taguchi2025ATel} show a blue continuum with narrow hydrogen emission lines ($v_{\rm{FWHM}}\approx900$\,km\,s$^{-1}$), gradually transitioning to exhibit redder continua over the eruption duration -- exhibiting narrower hydrogen and metallic lines such as Fe, Ca, before eventually developing broad absorption features due to TiO and VO molecules at late phases($>50$\,days since $g-$band peak, Section\, Figure \ref{fig:opt_spec}, \ref{sec:modeling_lrn_spec}). 

From October 2025 to February 2026, we obtained a series of optical spectra of the eruption, that reveal clear signs of shock interactions between the ejecta and circumstellar material (Section\,\ref{sec:modeling_lrn_spec}). The H$\alpha$ line in a spectrum obtained 4 days after the $g-$band peak shows an emission component and a narrow absorption feature whose minimum is blueshifted by $\approx15$\,km\,s$^{-1}$ (Figure \ref{fig:opt_spec}). During the next 80 days, the emission component gradually narrows and becomes stronger, indicating that the fast ejecta decelerate as they crash into the neutral ejecta and shock them (Figure \ref{fig:opt_spec}). The evolution of the velocity of the emission component matches the expectation for momentum conserving interactions between fast moving ejecta and circumstellar material with a density profile $\rho(r) \propto r^{-2}$ (Figure\,\ref{fig:opt_spec}). The line profile at late times is asymmetric, with a stronger blueshifted emission component from shocked ejecta along the line-of-sight. Such profiles have been observed in Type IIn supernovae and attributed to circumstellar material in disk geometries viewed along the equatorial plane \cite{Smith2015}. 

On 2025 December 24, 47 days after the \emph{g-}band peak, we obtained \emph{James Webb Space Telescope} (JWST) observations of the eruption, that coincided with the second sky survey of SPHEREx (Figure \ref{fig:evolution_properties}). Modeling the combined SPHEREx and JWST SEDs (Section\,\ref{sec:modeling_jwst}), we find a warmer stellar photosphere compared to the precursor -- surrounded by an optically thin dust shell located at a distance of $\approx5\times10^{4}$\,R$_{\odot}$. The inferred radius of the dust shell lies beyond both the maximum distance reachable by material ejected during the eruption or the seven-year precursor phase, and the maximum radius out to which dust could be sublimated by the 2025 eruption (Section~\ref{sec:disc_prog}). We therefore conclude that the \emph{JWST} observations probe remnants of dust around the progenitor that were formed during its prolonged phase of enhanced mass loss. Combining the properties of the dust shell inferred from the progenitor SED with its radial extent inferred from \emph{JWST} yields a total dust mass of $\approx6\times10^{-5}$\,M$_{\odot}$ surrounding the quiescent progenitor, about ten years before eruption (Section~\ref{sec:disc_prog}).  

The 2025 eruption of AT\,2025abao displays the defining hallmarks of Luminous Red Novae (LRNe; Figures \ref{fig:evolution_properties} and \ref{fig:opt_spec}) \cite{Tylenda11, Pastorello2019a, Blagorodnova2021}, a class of transients established to arise from impulsive energy injection into a common envelope during the merger of a star with its companion\cite{Tylenda11, Pastorello2019a}. However, AT\,2025abao is set apart from all previously known LRNe by its progenitor primary, which is unambiguously an evolved AGB star with an extended convective envelope. This progenitor system resembles those considered in the classical CEE formalism for compact binary formation \cite{Paczynski1976IAUS, Webbink84, Wilson2019, Wilson2020}, in which a loosely bound convective common envelope can theoretically be ejected by an inspiraling companion \cite{Klencki2021}. In contrast, all known LRN progenitors have primary stars on the main sequence or in the Hertzsprung Gap \cite{Blagorodnova2020, Pastorello2019a}, where the more tightly bound radiative envelopes are expected to inevitably drive the system toward merger \cite{MacLeod2022}. Thus, AT\,2025abao provides the first view of a classical CEE progenitor system continuously monitored from its onset through its culmination.

\begin{figure}[H]
    \centering
    \includegraphics[width=\textwidth]{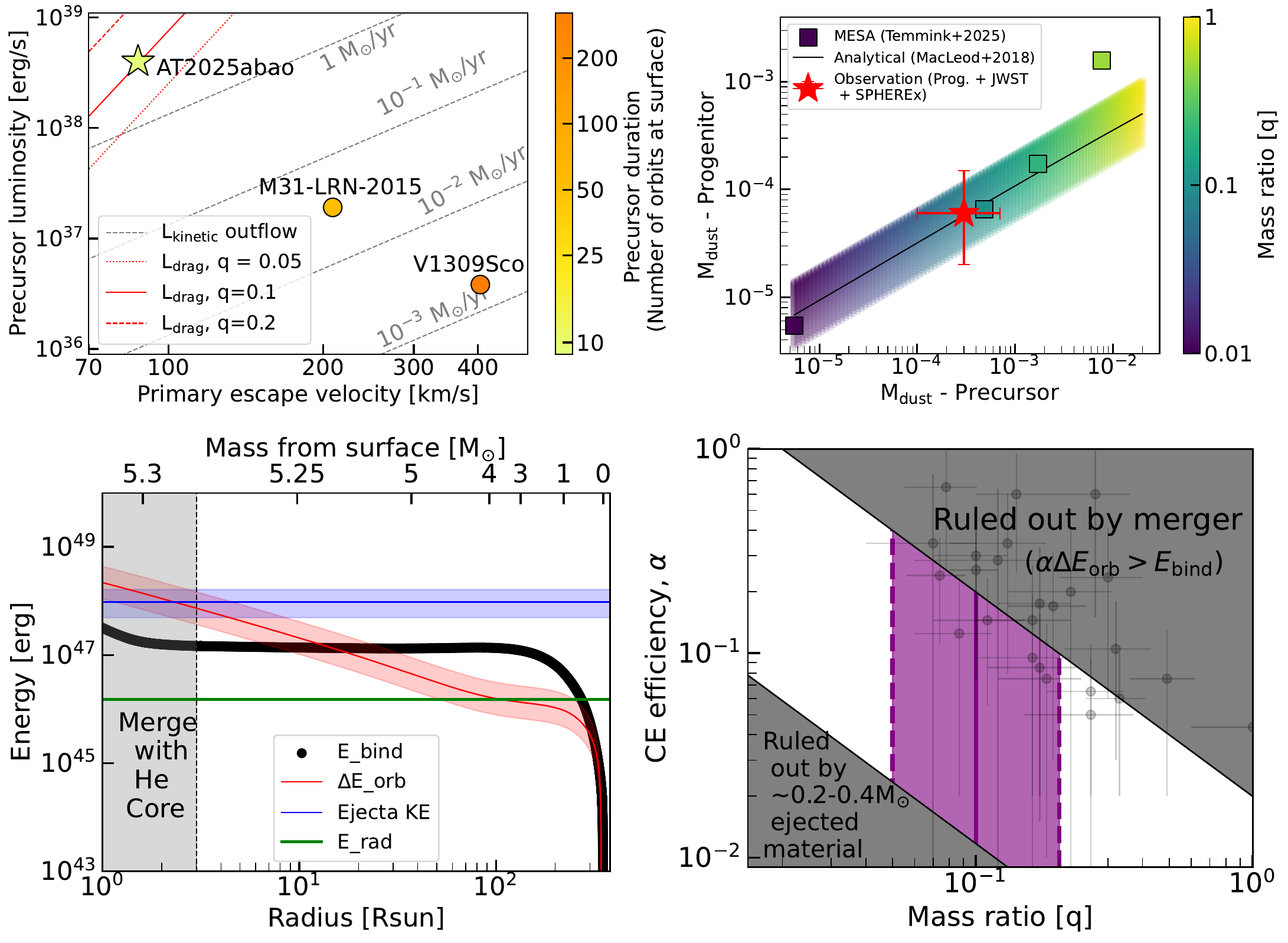}
    \caption{\textbf{Understanding CEE.}\\
    \small{\emph{Top Left:} Peak precursor luminosity versus primary escape speed for AT\,2025abao, V1309Sco \cite{Tylenda11}, and M31-LRN-2015 \cite{Blagorodnova2020}, colored by precursor duration in orbital periods at the primary's surface. Lines show the kinetic luminosity of outflows at various mass-loss rates (dashed gray, Equation \ref{eqn:lkin}) and drag luminosity from an inspiraling companion (red, Equation \ref{eqn:ldecay}) for the mass ratios in the legend. \\
    \emph{Top right:} Theoretical estimates of dust mass around the quiescent progenitor (black line: analytical, squares: MESA simulations) versus that ejected during the precursor (Section\,\ref{sec:disc_precursor}), colored by binary mass ratio. The shaded region spans the progenitor dust range by varying the analytical-model parameters (Section\,\ref{sec:disc_mass_loss_comparison}). The red star shows measured values with 1$\sigma$ uncertainties.  \\
    \emph{Bottom left: } Envelope binding energy versus radius (black) from the MESA model of the 7\,M$_{\odot}$, 340\,R$_{\odot}$ AGB primary star, inferred orbital energy dissipated by the companion (red), observed ejecta kinetic energy (blue), and observed radiated energy during the precursor (green). Solid lines are median values, shaded regions mark 1$\sigma$ uncertainties. \\
    \emph{Bottom Right:} CE efficiency $\alpha$ versus binary mass ratio. The inferred mass ratio and its 1$\sigma$ uncertainty is marked with the purple solid line and shaded region. Black shaded regions are ruled out by a core-merger in this system and its observed ejected mass. Black points show $\alpha$ measurements from post-CE binary systems \cite{deMarco2011} (Section \ref{sec:disc_alpha}).}}
    \label{fig:precursor_comparisons}
\end{figure}

We leverage our multi-wavelength dataset to draw the first direct constraints on the CEE phase, starting with the nature of the companion star. As the companion approaches the primary star, unstable mass transfer from the primary star over the centuries preceding CEE is expected to drive outflows of unbound material that enhance the mass loss rate of the quiescent system \cite{Tout1988, Mohamed2007, deVal-Borro2009,Chen2017}. Comparing the dust mass measured around the quiescent system to theoretical estimates (Section\,\ref{sec:disc_mass_loss_comparison}) from detailed simulations \cite{Temmink2023, Temmink2025} and analytical prescriptions \cite{MacLeod2018b,MacLeod2020} of unstable mass transfer constrains the mass ratio ($q=M_{\rm{secondary}}/M_{\rm{primary}}$) of the progenitor binary to $q\approx0.1$, indicating that the companion is a low mass $\approx0.7$\,M$_{\odot}$ star (Figure\,\ref{fig:precursor_comparisons}).  

Observations of previously known LRNe show gradual brightening by several magnitudes over tens to hundreds of orbits before the eruption \cite{Tylenda11, Blagorodnova2020}, attributed to outflows from the outer $L_2$ Lagrangian point as the companion approaches the primary \cite{Pejcha14, Pejcha2017ApJ}. These outflows form an expanding photosphere around the binary, preventing a direct view of the inspiral phase. However, the precursor phase of AT\,2025abao is challenging to explain in its entirety by the same mechanism -- the low escape speed of the AGB primary of AT\,2025abao makes such $L_2$ outflows energetically insufficient to power the observed luminosity ($\approx10^{38}$\,erg\,s$^{-1}$), even if all of the kinetic energy were radiated away (Figure \ref{fig:precursor_comparisons}). The long orbital period around the AGB primary also implies that the $\sim7$ year precursor phase spans fewer than ten orbits at the surface, substantially shorter than the other LRNe and that expected for dynamical outflows.

Instead, the precursor can be explained by energy deposition from a companion spiraling through the primary's outer envelope (Figure \ref{fig:precursor_comparisons}), signaling the onset of CEE. Adopting the Bondi-Hoyle formalism \cite{Bondi1944} (Section\,\ref{sec:disc_precursor}), the energy dissipation rate due to frictional drag on a $\approx0.7$\,M$_{\odot}$ companion orbiting near the photosphere of the AT\,2025abao primary is $L_{\rm drag}\approx3\times10^{38}$\,erg\,s$^{-1}$, consistent with the observed precursor excess luminosity. The corresponding timescale, $|E_{\rm orb}|/L_{\rm drag}\sim1000$ days, or a few orbits, is also broadly consistent with the observed duration of the precursor. The near-constant photospheric temperature and steadily increasing radius during the precursor phase (Figure \ref{fig:evolution_properties}) indicate frictional drag induces a quasi-static expansion of the AGB star -- likely mediated by efficient transport of excess energy to the stellar surface by convection \cite{Wilson2019, Wilson2020}. The total radiated energy during this phase is $\approx10^{46}$\,erg. 

The dust mass measured from the first SPHEREx observation constrains the mass outflow rate during this phase to $\sim10^{-2}$\,M$_{\odot}$\,yr$^{-1}$. Assuming this mass is ejected at the escape speed of the primary ($\approx85$\,km\,s$^{-1}$), the low outflow kinetic power ($\approx10^{37}$\,erg\,s$^{-1}$) compared to the observed luminosity ($\approx10^{38}$\,erg\,s$^{-1}$) indicates that majority of the dissipated orbital energy is radiated during this outer inspiral. We can independently estimate the mass of the companion by comparing the measured mass to theoretical expectations of mass loss due to an inspiraling companion (Section\,\ref{sec:disc_precursor}), which yields a mass ratio $q\approx0.1$, consistent with that inferred from dust around the quiescent progenitor. As the energy dissipation rate increases rapidly with decreasing orbital separation, the final 300 days of the precursor phase marked by a temperature drop likely mark the breakdown of quasi-static expansion and the onset of a rapid inspiral from the outer layers to the core. This period corresponds to about ten dynamical timescales of the primary star's envelope, in line with theoretical predictions of the inspiral duration in such systems \cite{Ivanova2013araa, Sand2020}.

The energetics of the 2025 LRN eruption constrain the mass of the ejecta and the circumbinary material (CBM; Section\,\ref{sec:disc_lrn_eruption}), as well as the fate of the companion star at the end of CEE. As the rise time of the eruption is governed by photon diffusion through the ejecta \cite{Padmanabhan2001}, our observations imply $M_{\rm{ej}}\approx0.1$--$0.2$\,M$_{\odot}$ ejected during the merger. The fast continuum photospheric expansion coincident with spectral line deceleration indicate that these ejecta expand freely in the polar direction but are obstructed by CBM in the equatorial plane \cite{Metzger2017, MacLeod17}. The observed velocity deceleration from $\approx900$\,km\,s$^{-1}$ to $\approx150$\,km\,s$^{-1}$ (Figure \ref{fig:opt_spec}) indicates that the ejecta sweep up $\sim$five times their mass. Assuming the CBM covers $\sim20$\% of the merger ejecta \cite{MacLeod2018b} gives a total CBM mass of $\sim0.1$--$0.2$\,M$_{\odot}$, broadly consistent with the CBM mass estimated from the SPHEREx SED. The kinetic energy dissipated by these interactions is $\approx3\times10^{47}$\,erg  while the energy radiated during the plateau phase $\approx4\times10^{46}$\,erg, indicating that shock-interactions release sufficient energy to power the observed light curve of the AT\,2025abao eruption, as previously theorized \cite{Metzger2017, Kirilov2025}.

The ejecta properties can be used to infer the fate of the binary (Section \ref{sec:disc_merger}). Assuming that the dissipated orbital energy is the primary energy reservoir, the observed kinetic energy of $\approx10^{48}$\,erg requires the companion to have inspiraled to a radius $\approx3$\,R$_{\odot}$ -- coincident with the He core of the primary, suggesting that the companion merged with the core. A merger is further supported by the low unbound mass of $0.2-0.4$\,M$_{\odot}$ that corresponds to only $\approx10\%$ of the primary's envelope. This indicates that the envelope is still intact, rather than being entirely ejected, as expected for a surviving binary. Finally, the AT\,2025abao eruption -- indicative of impulsive energy injection -- is extremely similar to other LRNe that are known stellar mergers. We therefore suggest that AT\,2025abao represents a CEE event that culminated in the merger of the binary components, which can be directly tested with future observations (Section \ref{sec:disc_merger}). 

During the rapid plunge to the primary core, the companion dissipated $\approx10^{48}$\,erg, of which only $\approx1$\% was radiated away ($\approx10^{46}$\,erg) and a similarly small amount ($\approx10^{46}$\,erg) was expended to unbind the observed ejecta (Section\,\ref{sec:disc_alpha}, Figure\,\ref{fig:precursor_comparisons}). Instead, almost all orbital energy was expended to accelerate the ejecta to high kinetic energy ($\approx10^{48}$\,erg). This is in contrast to the slow inspiral that powered the precursor phase, when most of the dissipated energy was radiated away. Our observations thus indicate a transition from a radiatively efficient slow inspiral to a kinetically efficient plunge, consistent with theoretical expectations of CEE inside a convective envelope \cite{Wilson2019, Wilson2020}. 

We examine AT\,2025abao in the context of the widely used energy formalism \cite{vandenHeuvel1976,Tutukov1979,Webbink84} (Section\,\ref{sec:disc_alpha}), which predicts the outcome of CEE in binary systems using an efficiency parameter $\alpha$ that relates the fraction of the dissipated orbital energy ($\Delta E_{\rm{orb}}$) to the envelope binding energy ($E_{\rm{bind}}$). For AT\,2025abao, an inspiral of the low-mass companion (q$\approx0.1_{-0.05}^{+0.10}$) down to the He core of the 7\,M$_{\odot}$ primary releases $\Delta E_{\rm{orb}}\approx8\times10^{47}$\,erg while the binding energy of the primary envelope is $\approx 1.6\times10^{47}$\,erg. A merger that fails to eject the envelope therefore implies $\alpha \lesssim 0.2_{-0.1}^{+0.2}$. The partial unbinding of the outer envelope requiring $\approx10^{46}$\,erg further implies $\alpha \gtrsim 0.01$, bounding the efficiency as $0.01 \lesssim \alpha \lesssim 0.4$. This constraint is consistent with that inferred from post-CE binaries, which suggest low values of $\alpha \sim0.1$-$0.3$ \cite{Zorotovic2010, Scherbak2023}. However, this formalism assumes that transferred orbital energy is primarily expended in gravitationally unbinding the envelope, whereas our observations reveal that the dominant energy sink was instead the acceleration of a small fraction of the envelope to speeds well exceeding the local escape speed. The formalism thus provides a useful parametric prescription for predicting CEE outcomes, but fails to capture all the underlying physics.

The inferred bipolar morphology of ejecta in AT\,2025abao -- expanding freely in the polar directions but confined by equatorial material -- confirms the long-standing hypothesis that bipolar planetary nebulae can be shaped by CEE in binary stars \cite{Livio1993, Soker1997, Ricker2012}. CEE in giant stars, like that probed here, is also a key step in forming close white dwarf binary systems \cite{Webbink84} (Section\,\ref{sec:disc_alpha}), including gravitational wave sources detectable by the future Laser Interferometer Space Antenna. Finally, the multi-decade observations of AT\,2025abao demonstrate the way forward to use long-baseline, multi-wavelength surveys to connect pre-CEE progenitors to their outcomes. Upcoming surveyors such as the Vera Rubin Observatory \cite{Ivezic2019} and the Nancy Grace Roman Space Telescope should routinely find these systems in $\approx 100$\,Milky-way like galaxies within $\approx 15$\,Mpc -- charting the landscape of binaries undergoing CEE. 

\newpage

\setcounter{firstbib}{\value{NAT@ctr}}

\setcounter{firstbib}{\value{NAT@ctr}}

\newpage

\newpage 

\section{Methods}
\subsection{Discovery}
\label{sec:discovery}
The slow brightening of the star WNTR23bzdiq at J2000 coordinates $\alpha=$00:38:48.60, $\delta=$+40:46:08.21, detected by the Wide-field Infrared Transient Explorer (WINTER, \cite{Frostig2025_winter}) survey was first reported in early 2025 by \cite{Karambelkar2025}. This source was independently identified in a dedicated search for transients using data from the NEOWISE \cite{Mainzer2014, Meisner2019} survey (survey name WTP19aalzlk), and was also detected in the optical by the Zwicky Transient Facility (\cite{Bellm2019, DeKany2020, Masci2019, Graham2019}, survey name ZTF21abcvhyx). Its rapid brightening in late-2025 was first noted on 2025 October 17 UT by K. Itagaki and independently by R. Fridrich on 2025 October 18 \footnote{http://www.cbat.eps.harvard.edu/unconf/\\followups/J00384865+4046079.html}. It was first reported to the Transient Name Server (TNS) and assigned the name AT\,2025abao by \cite{Kechin2025TNS} on 2025 October 20. 

On 2025 November 2, \cite{Taguchi2025ATel} noted its spectroscopic transition to hot temperatures and the emergence of hydrogen emission lines, resembling early-time properties of the class of eruptions called LRNe \cite{Pastorello2019a}. Over the next two months, the spectrum transitioned back to colder temperatures with molecular absorption features (Section\,\ref{sec:modeling_lrn_spec}) -- the hallmark observational signatures of LRNe \cite{Pastorello2019a}. This robustly establishes AT\,2025abao as an LRN, and rules out other low-luminosity eruptions such as Intermediate Luminosity Red Transients (ILRTs) \cite{Cai2021}. \cite{Reguitti2026} published optical-infrared observations of this source and an independent analysis of the LRN eruption. 

In our analysis, we assume a distance to M31 of 0.77\,Mpc \citep{Ferrarese2000}, and correct for a Galactic line-of-sight extinction of E(B$-$V) = 0.09\,mag \citep{Schlafly12}. Following \cite{Karambelkar2025}, we do not assume additional reddening from M31, as this source is located at the outskirts of the galaxy.

Before proceeding, we briefly clarify the terminology regarding LRNe and CEE adopted in this paper. As the late-2025 eruption AT\,2025abao is a bonafide member of the phenomenological LRN class, we refer to it as such throughout the article. LRNe have been established to originate from stellar mergers in binary systems \cite{Pastorello2019a}. AT\,2025abao is the first known LRN in which the progenitor primary star is an evolved AGB star with a convective envelope. This makes it a unique probe of classical CEE, which we explore in this paper.

\subsection{Follow-up Observations}
\label{sec:followup_obs}
\subsubsection{Optical and near-IR photometry from ground-based surveys}
\cite{Karambelkar2025} presented optical-infrared photometric observations from 2005 to April 2025, compiled using data from the Canada France Hawaii Telescope (CFHT), the Palomar Transient Factory (PTF, \cite{Law09}), the intermediate Palomar Transient Factory (iPTF \cite{Cao16}), the PanSTARRS (PS1, \cite{Chambers16, Magnier13, Flewelling2020}), ZTF, WINTER, and NEOWISE surveys. We now describe additional observations of this source. 

In the optical, ZTF continued observing AT\,2025abao as part of its routine survey operations in the \emph{g}, \emph{r} and \emph{i} bands from April 2025 to February 2026. The ZTF photometry was retrieved using the Skyportal/Fritz alert brokering system \cite{vanderWalt2019, Coughlin2023}. Additionally, we observed it with the Spectral Energy Distribution Machine (SEDM, \cite{Blagorodnova2018}) camera on the 60-inch telescope at Palomar Observatory in the \emph{ugri} filters on several epochs between November 2025 and February 2026, and reduced the data using \texttt{FPIPE} \cite{Fremling16}. We also monitored it with the GROWTH-India Telescope (GIT, \cite{Kumar2022_git}) in the SDSS $g^\prime$, $r^\prime$, $i^\prime$, and $z^\prime$ filters at approximately daily cadence and reduced the data with the GIT data processing pipeline \cite{Kumar2022_git}. Finally, we downloaded archival red (RG610) and infrared (I) band images from the second Palomar Optical Sky Survey (\cite{Reid1991}, POSS-2) taken on 1989 October 1 and 1993 September 23, in which the progenitor source of AT\,2025abao is clearly detected. We calibrated the images relative to the PS1 catalog and performed aperture photometry to determine the archival brightness of the source.

In the near-infrared, the source was observed as part of WINTER's routine survey from 2025 May to 2025 September during its precursor phase. After the rapid brightening, we obtained Y, J, and Hs band observations with WINTER, at a cadence of $\approx3$ days until February 2026. These data were processed using the WINTER data reduction pipeline implemented using \texttt{mirar} \footnote{\url{https://github.com/winter-telescope/mirar}} \cite{mirar_zenodo}. Additionally, we observed the source with the Wide-Field InfraRed Camera (WIRC) \cite{Wilson2003} on the 200-inch Hale Telescope at Palomar Observatory in the J, H, and Ks bands on 2025 December 26; these observations were processed using python routines for reduction and calibration.

\subsubsection{High cadenced optical observations}
We observed AT\,2025abao with the prototype \textit{Cerberus} high-speed imager mounted at the prime focus of the 200-inch Hale telescope at Palomar Observatory \cite{cerberus_mo_inprep}, to search for any evidence of rapid variability.The observations were performed on 2025-11-25, lasting 14.7\,min in the SDSS-\textit{r} band at 1\,s cadence. AT \, 2025abao was clearly detected with the signal-to-noise ratio (S/N) $>300$ in individual exposures.

\subsubsection{Swift ultraviolet and X-ray observations}
We triggered target-of-opportunity (ToO) observations (PI : Karambelkar) with the Neil Gehrels Swift Observatory \cite{Gehrels2004} to obtain ultraviolet (UV) and X-ray imaging on 2025 November 3. The source was detected in all UV bands (u, UVW1, UVM2, UVW2) by the UltraViolet/Optical Telescope (UVOT, \cite{Roming2005}) but no X-rays were detected in the X-ray Telescope (XRT; \cite{Burrows2005}) observations. We continued monitoring the source with Swift/UVOT from 2025 November 8 to November 15 with a cadence of $\approx$2 days. 

The UVOT observations were downloaded and reduced using \texttt{uvotredux}\footnote{\url{https://github.com/robertdstein/uvotredux}} \cite{uvotreduxv0.3.1}, a Dockerized open-source python package for automated UVOT data reduction using  \texttt{swifttools}\footnote{\url{https://github.com/Swift-SOT/swifttools}} and \texttt{HEASoft}\footnote{\url{https://heasarc.gsfc.nasa.gov/docs/software/lheasoft/}}. Photometry was extracted assuming a $5''$ aperture centred on the position of the transient, and background was estimated using an offset $10''$ aperture. The XRT observations were reduced using the online tool \cite{Evans2009}.

\subsubsection{SPHEREx spectrophotometry}
The field of AT\,2025abao was first observed by the SPHEREx space mission from 0.7-5\um as part of its survey when the source was in its precursor phase, between 2025 July 9 to 2025 August 15, with 95\% of the observations conducted before 2025 July 22. The field was observed again between 2025 December 18 to 2026 January 1, during its eruption phase. We used the public SPHEREx spectrophotometry tool to obtain the 0.7-5\,\um spectrum of this source at both epochs. The data were examined for quality cuts similar to the metrics described in \cite{Karambelkar2026_spherex}.

\subsubsection{Ground-based optical and near-IR spectroscopy}
\cite{Karambelkar2025} presented optical and near IR spectra of the precursor phase obtained between 2023 October to 2025 January. Here, we describe further spectroscopic observations.

On the night of 2025 September 9, when the source was still in its precursor phase, we obtained a medium resolution (R$\approx2700$) near-IR spectrum with the SpeX spectrograph on the NASA Infrared Telescope Facility (IRTF) as part of program 2025B061 (PI: Frostig). The spectrum was reduced using the \texttt{idl} package \texttt{spextool} \citep{Cushing2004} and corrected for telluric absorption using the package \texttt{xtellcor} \citep{Vacca2003}.

During the 2025 eruption, we obtained a series of R$\approx$8800 optical spectra with the Gemini Multi-Object Spectrograph (GMOS) on the 8-m Gemini North telescope on several nights from 2025 November 11 to 2026 February 6. These observations were obtained as part of Director's Discretionary Time Program GN-2025B-DD-111 (PI: Karambelkar), and comprised six spectra using the R831 grating (R$\approx8800$) and six with the B1200 grating (R$\approx7400$), all using the 0.25\,arcsecond slit. Data were processed using Data Reduction for Astronomy from Gemini Observatory North and South (DRAGONS, \cite{Labrie2019}) software. 

We obtained a series of five R$\approx6000$ optical spectra with the Next Generation Palomar Spectrograph (NGPS) on the 200-inch Hale Telescope at Palomar Observatory, from 2025 November 11 to 2026 January 26 (PI: Kasliwal). These spectra utilized the 0.4\,arcsecond slicer providing the highest spectroscopic resolution. The data were reduced using standard procedures for long-slit spectroscopic data reduction and flux calibration using a custom pipeline developed for NGPS. 

We obtained a R$\approx4000$ optical spectrum with the Binospec spectrograph \citep{Fabricant+2019} on the 6.5\,m MMT telescope on 2025 November 13 (PI: Rehemtulla). The 600 lines/mm grating was used in 3x120s exposures centered at 6500 \AA~and 5x120s exposures centered at 8500 \AA. The data are reduced with standard procedures implemented in \texttt{PypeIt} \citep{pypeit}, and the resulting 1D spectra are trimmed to remove areas of the spectrum which are contaminated by light from the second order of the grating.

We obtained a R$\approx25000$ optical spectrum with the FIbre-fed Echelle Spectrograph (FIES, \cite{Telting2014}) mounted on the Nordic Optical Telescope (NOT) on 2025 November 23 through Fast-Track proposal 72-405 (PI: de Wet). Observations consisted of $2\times1800$~s exposures in low resolution mode, and were reduced using the standard FIEStool\footnote{\url{https://www.not.iac.es/instruments/fies/fiestool/}} pipeline.

We obtained two R$\approx$2500 optical spectra with the OSIRIS+ spectrograph on the 10.4-m Gran Telescopio Canarias (GTC) on the nights of 2025 December 25 and 2026 January 20 under program 24-GTC20-A/25B (PI : Jones).  On both occasions, three spectra each of 60s exposure time were obtained using the R2500V and R2500R gratings, in combination with a 0.6\,arcsecond longslit. Data were reduced and coadded using \texttt{PypeIt} \cite{pypeit}.

We obtained a series of R$\approx650$ spectra from 2025 October 24 to 2026 January 30 with the Hanle Faint Object Spectrograph (HFOSC) on the 2.4\,m Himalayan Chandra Telescope using gratings Gr\#7 and Gr\#8 (programs HCT-2025-C3-P27, HCT-2026-C1-P13, PI: Chamoli). The spectra were reduced using standard procedures in IRAF.

We obtained a spectrum with the Kast Double Spectrograph on the at the 3.05 meter C. Donald Shane telescope at the Lick observatory (PI: Rich) on the night of 21 November 2025. On the red side detector, the 300/7500 grating was used, which has a dispersion of 2.55 \AA/pixel, and $R\approx2800$. On the blue side, the 600/4310 grism was used, which has a dispersion of 1.02 \AA/pixel, with $R\approx3850$. The spectra were reduced using the UCSC Spectral Pipeline\footnote{\url{https://github.com/msiebert1/UCSC\_spectral\_pipeline}}, \cite{Siebert20}, based on methods described by \cite{Foley03, Silverman2012}.

We obtained three R$\approx100$ spectra using the Integral-Field Unit spectrograph Spectral Energy Distribution Machine (SEDM, \cite{Blagorodnova2018, Kim2022}) on the 60-inch telescope at Palomar Observatory on 2025 November 4, 5, and 13. These spectra were reduced using the \texttt{pysedm} software \cite{Rigault2019}.

We obtained two R$\approx3000$ NIR spectra of the eruption with the Near-InfraRed Echelle Spectrograph (NIRES) spectrograph on the Keck II telescope on 2025 November 9 and 2025 December 12 (PI: Kasliwal). We also obtained a spectrum with the Triplespec spectrograph on the 200-inch Hale Telescope at Palomar Observatory on 2025 December 8, and one with NASA/IRTF on 2025 December 24. These spectra were reduced using \texttt{spextool} and \texttt{xtellcor}, as described above. 

\subsubsection{James Webb Space Telescope mid-infrared observations}
We obtained mid-infrared observations with the \emph{James Webb Space Telescope} on 2025 December 24 ---  under Director's Discretionary Time Program 12487 (PI: Karambelkar). AT\,2025abao was spectroscopically observed with the Medium Resolution Spectrograph (MRS) on the Mid InfraRed Instrument (MIRI, \cite{Wright2023, Dicken2024}), and imaged with MIRI in the F1800W, F2100W, and F2550W filters. We downloaded standard Level 3 data products available on the Mikulski Archive for Space Telescopes (MAST). The source is detected with S/N$\geq100$ in all \emph{JWST} observations. We also obtained a GMOS optical spectrum, SpeX NIR spectrum, and WIRC NIR photometry within a day of the \emph{JWST} observations. The serendipitous SPHEREx observations were conducted within a week around the epoch of the \emph{JWST} observations.

\subsection{Spectral energy distribution evolution}
\subsubsection{Modeling the progenitor SED}
\label{sec:modeling_progenitor}
\cite{Karambelkar2025} modeled the archival spectral energy distribution (SED) of the progenitor star and derived T$_{\rm{eff}}=3500^{+200}_{-200}$\,K, log(L/L$_{\odot}$)=4.2$\pm$0.1, R$_{\rm star}\approx350$\,R$_{\odot}$ consistent with a 7$\pm$2\,M$_{\odot}$ star in its early AGB phase. The properties of the dust shell depended on the method used : modeling with the PHOENIX model photospheres \cite{Husser2013} and the radiative transfer code \texttt{DUSTY} \cite{Ivezic97} gave T$_{\rm{dust}}\approx1450$\,K, and $\tau_V\approx2.7$ while a comparison to pre-computed Grid of Red supergiant and Asymptotic Giant Branch ModelS (GRAMS; \cite{Sargent2010,Sargent2011,Srinivasan2011}) suggested T$_{\rm{dust}}\approx1000$\,K and $\tau_{V}\approx1.9$. \cite{Karambelkar2025} adopted dust opacities from \cite{Draine84} for their \texttt{DUSTY} modeling. However, our \emph{JWST} data indicate that the dust is comprised of warm oxygen-rich silicates (see Section \ref{sec:modeling_jwst}). Using the updated opacities, we redo the \texttt{DUSTY} progenitor SED modeling using Markov Chain Monte Carlo (MCMC) sampling implemented using the python package \texttt{emcee} \cite{Mackey-Foreman2013} and the methods outlined in \cite{Karambelkar2025}. Briefly, we used the log\_g=0.0 PHOENIX models as inputs to \texttt{DUSTY} to generate model SEDs for given values of stellar effective temperature ($T_{\rm{star}}$), dust shell temperature ($T_{\rm{dust}}$), and dust optical depth ($\tau_{V}$), which are free parameters in our modeling. Similar to \cite{Karambelkar2025}, we use a standard MRN distribution for the dust grain size and a dust radial density profile $\rho(r) \propto r^{-2}$ expected for wind-like mass loss. We also assume the dust outer to inner shell radius ratio (Y) of 2 for these models, but we note that the constraints on $T_{\rm{star}}$, $T_{\rm{dust}}$, and $\tau_{V}$ are not sensitive to the exact values of this parameter. We fit the models to the observed SED to derive posterior distributions of these parameters, that are shown in Extended Data Figure \ref{fig:corner_progenitor}.

From our updated modeling, we derive slightly higher stellar parameters of T$_{\rm{eff}}=3650_{-50}^{+50}$\,K, log(L/L$_{\odot}$) = 4.25$_{-0.02}^{+0.02}$, R$_{\rm{star}}=330\pm50$\,R$_{\odot}$, and dust parameters of T$_{\rm{dust}}=1100_{-100}^{+100}$\,K, and $\tau_{V}=1.9_{-0.3}^{+0.3}$ --- in good agreement with the GRAMS dust parameters. The inner radius of this dust shell is R$_{\rm{dust, in}}=4100_{-1000}^{+1000}$\,R$_{\odot}$. Comparing to solar abundance stellar evolutionary tracks from MESA Isochrones and Stellar Tracks (MISTv2.5; \cite{Dotter2016, Choi2016, Dotter2026})\footnote{downloaded from https://waps.cfa.harvard.edu/MIST/} and the PAdova TRieste Stellar Evolution Code (PARSEC v2.0 \cite{Costa2019a, Costa2019b, Nguyen2022, Nguyen2025}), we find these values consistent with those expected for a 7$_{-1}^{+1}$\,M$_{\odot}$ star in its early AGB phase, consistent with \cite{Karambelkar2025}. 

\subsubsection{Modeling the multiband SED during the slow precursor brightening}
\label{sec:modeling_precursor_sed}
\cite{Karambelkar2025} showed that the progenitor brightness did not vary substantially from 2005 to 2015. The even older POSS-2 measurements from 1989 presented here (Figure \ref{fig:full_lightcurve}) suggest that there were no substantial variations for at least a 26 year period. From early 2018 to October 2025, the source steadily brightened at all wavelengths. It first underwent an $\approx$1\,mag bump from 2020--2023. After this, the source rebrightened until December 2024 and then faded by 1 mag over the next 250 days in the \emph{g} and \emph{r} bands, however, staying constant in the \emph{i} band, and \emph{brightening} in the IR (\emph{J}, \emph{W1}, \emph{W2}) bands. It then remained in its low optical state for $\approx 50$ days until October 2025, when it underwent its dramatic eruption.
 
We constructed multiwavelength SEDs by combining ZTF, NEOWISE, and SPHEREx observations during this phase. We interpolated between the measurements using Gaussian Process (GP) regression to fill in temporal gaps. The GP model, implemented with the python package \texttt{scikit-learn} \citep{scikit-learn}, employed a Radial Basis Function kernel with an added White Noise kernel to account for observational noise, similar to \cite{Karambelkar2025}. For epochs between 2018 to 2024, we fit the SEDs using PHOENIX photospheric models, \texttt{DUSTY}, and \texttt{emcee} as described in Section \ref{sec:modeling_progenitor}. The resulting luminosities, stellar radii, and temperatures are plotted as solid lines with the uncertainties marked with shaded regions in Figure \ref{fig:evolution_properties}. We note that during this phase, the SEDs only have five points covering the 0.4 to 4.5\um wavelength range, of which several \emph{g-}band measurements are low signal-to-noise. Consequently, the inferred values have larger uncertainties compared to the progenitor SED (e.g., the uncertainty on stellar temperature range from $\approx100-200$\,K during this phase compared to the $\approx50$\,K uncertainty on the progenitor temperature). Despite this, we do not find any changes in the stellar temperature larger than $\approx\pm200$\,K during this phase. This is also consistent with the \emph{g-r} and \emph{r-i} colors during this phase that do not show any variations larger than $\approx0.5$\,mag \cite{Karambelkar2025}. From early to mid-2025, we repeated the above modeling procedure, but for W1 and W2 fluxes we used the GP-interpolated values between the last NEOWISE epoch from 2024 and the first SPHEREx epoch from mid-2025. For the last $\approx20$ days of the precursor phase, we only have \emph{g} and \emph{r}-band observations, and \texttt{DUSTY} modeling is infeasible with just two data points. Instead, we simply fit a blackbody model to the SED; however, blackbody fits to only optical data can underestimate the luminosities and radii at these temperatures where the radiation is expected to peak in the infrared. Based on the 2019--2024 data, we find that only using \emph{gr} data underestimates the luminosity by a factor of $\approx2.3$ compared to modeling the full SED. For the final phase, we corrected for the luminosities and radii by this factor. These values are plotted as hollow circles in Figure \ref{fig:evolution_properties}, and should be interpreted with caution.

Our modeling shows that for most of the precursor phase except the final three hundred days, the star stayed at roughly constant temperature $T_{\rm{star}}\approx3800_{-200}^{+200}$\,K (possibly slightly warmer than the progenitor, as noted in \cite{Karambelkar2025} and \cite{Reguitti2026}), as it steadily expanded and became more luminous. The luminosity and radius increase occurs in two ``ramp-up" episodes (Figure \ref{fig:evolution_properties}), corresponding to the two bumps in the optical lightcurve. Both bumps are accompanied by small increases in dust optical depth (from $\tau_V\approx1.9$ to $\approx2.8$), evidenced by the IR brightness continuing to rise even after the optical fades. Additionally, the rapid optical decline at the end of the second bump is accompanied by a temperature drop to $\approx3400$\,K and a rapid $\approx1.5\times$ increase in luminosity and radius in two hundred days --- the fastest expansion rate during this phase. Towards the end of the precursor phase, our modeling shows that the star has brightened to about six times its initial luminosity, expanded by about 2.5 times its initial size, and cooled slightly compared to its initial temperature. The inner dust shell temperature and radii are poorly constrained due to the sparse IR SED coverage --- the temperature $T_{\rm{dust}}\approx1000_{-300}^{+300}$\,K while the radius expands slowly by a factor of 3 compared to the progenitor.

\subsubsection{SPHEREx pre-outburst SED modeling}
\label{sec:modeling_spherex}
The 0.7--5\,\um SPHEREx spectrum taken $\approx 120$ days before the eruption \emph{g-}band peak provides the best constraints on the pre-eruption state of the system. We combine this with contemporaneous \emph{gr} measurements from ZTF and model the SED using the PHOENIX models, \texttt{DUSTY}, and \texttt{emcee} as described in Section \ref{sec:modeling_progenitor}. We derive T$_{\rm{star}}=3460_{-60}^{+70}$\,K, T$_{\rm{dust}}=870_{-200}^{+200}$\,K, and $\tau_V=2.6^{+0.7}_{-0.3}$. The total integrated luminosity is L$=9.7_{-0.5}^{+1.1}\times10^{4}$\,L$_{\odot}$, corresponding to a dust shell inner radius $r_{dust,in}=1.6^{+1.4}_{-0.7}\times10^4$\,R$_{\odot}$ and a stellar effective radius $R_{\rm{star}}=860_{-30}^{+50}$\,R$_{\odot}$. The best-fit model to the SPHEREx data is shown in Figure \ref{fig:evolution_properties} and the posterior distributions on the parameters are shown in Extended Data Figure \ref{fig:corner_spherex}. We note that the SPHEREx spectrum exhibits NIR water vapor absorption bands stronger than that predicted by the best-fit photospheric model. The SED also exhibits a mixed-temperature behavior, wherein fitting only the NIR data ($\geq1$\,\um) yields a lower temperature (T$\approx3000$\,K) than the full optical+NIR data. Both these features can be attributed to increased molecular absorption from circumstellar material around the progenitor primary, as noted by \cite{Karambelkar2025} based on other spectra obtained during the precursor phase. 

We note that \cite{Reguitti2026} compared the SPHEREx SED with dusty stellar models and found good agreement of the 1-2.5\,\um region with $T_{\rm{eff}}\approx3100$\,K models -- in line with our lower derived photospheric temperatures from the NIR portion described above. However, we also note that their comparisons use stellar models with log\_g=1.0, while the observed radius during the precursor and the quiescent phase indicate log\_g$\approx0$.

\subsubsection{Photometric evolution of the LRN eruption}
\label{sec:modeling_lrn_phot}
After the precursor phase ended in September 2025, the source began brightening rapidly. We constructed multiwavelength SEDs using broadband UV to NIR photometric data obtained during this phase. We fit blackbody functions to the SEDs to derive photospheric radii, temperatures and luminosities, which are plotted in Figure \ref{fig:evolution_properties}. Based on its evolution, this eruption can be divided into three phases -- an initial rise lasting $\approx25$ days, a plateau-phase lasting $\approx80$ days, and a late decline lasting at least $\approx30$ days. 

The initial rise began when the ZTF \emph{g}-band started brightening on 2025 September 19 (MJD 60937), while the \emph{r}-band began brightening a bit later on 2025 October 2 (MJD 60950). In the next 25 days, the source brightened by 5 magnitudes in the \emph{g} and \emph{r} bands while simultaneously becoming hotter. During this phase, the temperature increased rapidly to $\approx8000$\,K compared to 3500\,K in the precursor, and the luminosity increased to $\approx2.5\times10^{39}$\,erg\,s$^{-1}$ --- about ten-times that at the end of the precursor phase. The photospheric radius at the start of the initial rise is about three times smaller than in the precursor phase, indicating that the photosphere has moved inward. During the rise, the photosphere expands rapidly with expansion speeds of $\approx330$\,km\,s$^{-1}$, significantly higher than the slow expansion in the precursor phase. 

On 2025 October 27 (MJD 60975), the rapid brightening stopped and the eruption transitioned to a ``plateau" phase. The luminosity increased steadily to $\approx6\times10^{39}$\,erg\,s$^{-1}$ over the next $\approx80$ days, during which the temperature dropped to $\approx3000$\,K. The \emph{g} band brightness peaked on 2025 November 7 (MJD 60986), after which it gradually faded by $\approx1$mag in the next 80 days. During this plateau phase, the bluer bands (\emph{u}, \emph{UVW1}, \emph{UVM1}, \emph{UVW2}) faded, with the bluest bands fading fastest, while the redder bands (\emph{r, i, J, Hs}) brightened with the reddest bands increasing fastest. The photospheric radius continued to expand with a speed of $\approx330$\,km\,s$^{-1}$. 

The plateau-phase ended on 2026 January 13, when the luminosity and photospheric radii started decreasing. As Figure \ref{fig:full_lightcurve} shows, the redder bands started to fade before the bluer ones, suggesting that the photospheric temperature increased slightly. The source was still in this phase by mid-February 2026, when it went into solar conjunction.

The peak luminosity of $\sim$few$\times10^{39}$\,erg\,s$^{-1}$, rapid transition from hot to cold temperatures and photospheric expansion speeds of $\approx330$\,km\,s$^{-1}$ are hallmark signatures of LRNe. Figure \ref{fig:evolution_properties} compares these properties with M31-LRN-2015 \cite{Williams2015, Kurtenkov15, MacLeod17} and V838Mon \cite{Munari2002AA}, which had similar peak luminosities as AT\,2025abao. The main difference is that the duration of the plateau phase in AT\,2025abao is longer than the other events. Other than this, AT\,2025abao is strikingly similar to these events.

\subsubsection{Spectroscopic evolution of the LRN eruption}
\label{sec:modeling_lrn_spec}
Extended Data Figure \ref{fig:all_optical_spectra} shows the optical and NIR spectroscopic evolution of AT\,2025abao respectively. Spectra presented by \cite{Karambelkar2025} spanning 2023 to early 2025, during the precursor phase, resembled late M-type stars with oxygen-rich molecules \cite{Karambelkar2025}. We present an additional NIR spectrum from IRTF during this phase, obtained on September 9, 2025 (Figure \ref{fig:all_nir_spectra}), that also exhibits these features. 

During the initial rise in October 2025, the spectrum had transitioned to a hot blue continuum with strong hydrogen emission lines, with gaussian full width half maximum (FWHM) of $\approx900$\,km\,s$^{-1}$. Several narrow metallic absorption lines such as Na~ID, Fe~II are also visible. In the next three months through the plateau and late decline phase, the spectrum gradually cooled down and developed absorption features of oxygen-rich molecules. This spectral evolution is the hallmark of LRNe (e.g., \cite{Blagorodnova2020, Pastorello2019a, Pastorello2021b}, Figure \ref{fig:opt_spec}), cementing the classification of AT\,2025abao.

Our spectroscopic coverage allows us to trace ejecta velocities in all phases of the eruption. Low-resolution spectra during the initial rise suggest that the FWHM velocities decreased from $\approx900$ to $600$\,km\,s$^{-1}$ until the \emph{g}-band peak. The subsequent higher resolution spectra from NGPS, GMOS, FIES, and Binospec show that this velocity decreased within twenty days to $\approx150$\,km\,s$^{-1}$ and remained low for at least eighty days post the $\emph{g}$-band peak (Figure \ref{fig:opt_spec}). 

The H$\alpha$ lines in these spectra exhibit a two-component profile, with a broader emission component and a narrower absorption component (Figure \ref{fig:opt_spec}).  The center of the absorption component in the earliest spectrum is $\approx20$\,km\,s$^{-1}$ blueshifted relative to rest wavelength, confirming the presence of slow moving, neutral circumbinary material (CBM). With time, the absorption component weakens while the emission component strengthens, indicating that the CBM is being swept up by the shock. At late times, the line profile is asymmetric, with a stronger blueshifted emission, indicative of a recombination front moving towards us along the line of sight. These spectra unambiguously show the presence of shock interactions between fast-moving merger ejecta and slower-moving material. Similar asymmetric line-profiles seen in interacting Type IIn supernovae have been attributed to disk-like geometries for the circumstellar material \cite{Smith2015}. This is in line with theoretical expectations for equatorially focused CBM geometry \cite{MacLeod2018b, Metzger2017}. Additionally, it provides a natural explanation for the fast photospheric expansion speeds despite the decelerating spectral line velocities --- the merger ejecta are mostly spherically symmetric and expand freely in the polar directions, setting the continuum radiation, while a small fraction are obstructed in the equatorial plane by CBM and decelerated via shock interaction. Our observations of the decelerating ejecta suggest that our view to AT\,2025abao is close to edge-on, consistent with that suggested by \cite{Reguitti2026} from a similar analysis of hydrogen line spectral profiles. We note that this exact picture has been theoretically proposed for LRNe by \cite{Metzger2017}, and our observations provide an unambiguous confirmation of this model.  

The NIR spectra during the plateau phase (Extended Data Figure \ref{fig:all_nir_spectra}) exhibit a bluer continuum compared to the precursor phase spectra, show several metallic absorption lines as well as absorption bands due to CO. Weak CO bands are present even in the earliest post-eruption spectrum taken on 2025 November 9, when the optical spectra do not exhibit any molecular absorption, indicating that some of the pre-existing CO gas remained intact even after the LRN eruption destroyed a large fraction. 

\subsubsection{Post-outburst JWST and SPHEREx SED modeling}
\label{sec:modeling_jwst}
Our \emph{JWST} observations taken 47 days after the \emph{g}-band peak, together with serendipitous SPHEREx and contemporaneous ground-based optical-NIR observations, provide the earliest optical to mid-IR SED of a post-eruption LRN. Figure \ref{fig:evolution_properties} shows the 0.5--25\um SED of the source. These observations show that the SED peaks around $\approx1$\,\um and declines at longer wavelengths. The SPHEREx spectrum at this phase differs substantially from that in the precursor phase, exhibiting much weaker molecular absorption bands and a bluer continuum. The \emph{JWST} MRS spectrum shows a clear, broad emission feature spanning 8--12\,\um, attributed to optically thin silicate dust. This emission feature exhibits a peculiar morphology, with humps on the red and blue side relative to the peak. 

We model the SED using \texttt{DUSTY} and \texttt{emcee}. For the \texttt{DUSTY} modeling, we assume the input radiation is a single blackbody that is surrounded by a dust shell comprising single-sized dust grains and a $\rho \propto r^{-2}$ density profile. We examined several dust compositions and find that the shape of the silicate emission feature can be best replicated using opacities for a combination of warm oxygen-rich silicates (\texttt{Sil-OW}) from \cite{Ossenkopf1992} and amorphous silicates of enstatite composition (\texttt{enst}) from \cite{Day1979}. Both these compositions have been commonly observed around evolved AGB and RSG stars (e.g., \cite{Verhoelst2009, Laywell_2026}). The blackbody temperature ($T_{\rm star}$), dust shell temperature ($T_{\rm dust}$), dust optical depth ($\tau_{V}$), dust grain size ($a$), dust shell thickness ($Y$), and the fractional abundance of \texttt{Sil-OW} relative to \texttt{enst} ($a_{\rm silow}$) are free parameters. We fit the models to the SED using MCMC implemented with \texttt{emcee} with broad, uniform priors and derive posterior distributions on the parameters, shown in Extended Data Figure \ref{fig:corner_jwst}. 

From our modeling, we derive T$_{\rm{star}}=3840_{-50}^{+50}$\,K, T$_{\rm{dust}}=870_{-120}^{+80}$\,K,  $\tau_{V}=0.10_{-0.04}^{+0.03}$, $a=0.11_{-0.03}^{+0.03}$\,\um, Y=$24_{-12}^{+16}$,  $a_{\rm silow}=0.6_{-0.1}^{+0.1}$ (uncertainties represent 16$^{th}$ and 84$^{th}$ percentiles). The total integrated luminosity L=$9.6_{-0.4}^{+0.4}\times10^{5}$\,L$_{\odot}$, and the radius of the inner dust shell r$_{\rm{dust, in}}=4.7_{-0.8}^{+1.6}\times10^{4}$\,R$_{\odot}$. 

\subsubsection{Periodicity searches in the lightcurve}
We analyze the optical lightcurve of AT\,2025abao using a Lomb Scargle (LS) periodogram \cite{Lomb76, Scargle82} to search for periodic signatures. No evidence of periodicity was found in the PTF/iPTF data from 2010 to 2015 (also noted in \cite{Karambelkar2025}). For the precursor phase, GP fits were used to detrend the smooth long-duration ($\geq500$\,days) variations out of the lightcurve \cite{Karambelkar2025}, and an LS analysis on the detrended data did not reveal any significant periodic signatures between 2018 to 2025. However, two consecutive $\approx87$-day duration bumps were seen in the \emph{r-}band lightcurve in 2023 \cite{Karambelkar2025}. The \emph{r-}band eruption lightcurve during the plateau phase exhibits low amplitude (few $\approx0.1$\,mag) flickering with timescales of a few days; however, we do not detect any significant periodicity in these data. The typical cadence of our observations rules out any periods longer than $\approx1$\,day during the plateau phase of the eruption.

We also analyze the high (1 second) cadenced observations taken during the eruption plateau using Cerberus, to search for short-timescale intrinsic variations. After correcting for common atmospheric variations using a collection of comparison stars, we find a normalized excess variance \citep[NXV;]{1997ApJ...476...70N, 1999ApJ...524..667T} of $3 \times 10^{-5}$, with lightcurve scatter of about 0.7\%.
This is consistent with the NXVs of the comparison stars, indicating no significant intrinsic variability in AT\,2025abao beyond the photometric noise.
However, we cannot rule out variability on longer timescales or outside the observation window, especially given the limited (14.7 minutes) baseline of these observations.

\subsection{Understanding Common Envelope Evolution}
\subsubsection{Enhanced mass loss around the quiescent progenitor system}
\label{sec:disc_prog}
Our archival photometric observations show no substantial variations from 1989 to 2018. The progenitor SED constructed using data from 2010 to 2012 thus represents the state of the system in the three decades prior to eruption.

Our \texttt{DUSTY} progenitor modeling can be used to derive the mass loss rate of the progenitor primary, using 
\begin{equation}
    \dot{M} = \eta_{gd} \frac{4\pi v_{\rm{wind}} R_{\rm{in}} \tau_{V}}{\kappa_{V}} 
\end{equation}
where $\eta_{gd}$ is the gas-to-dust ratio, v$_{wind}$ is the terminal wind velocity, $\tau_{V}$ is the dust optical depth and $\kappa_{V}\approx5\times10^3$\,cm$^{2}$\,g$^{-1}$ is the opacity in the V-band. Assuming a typical v$_{wind}\sim15-20$\,km\,s$^{-1}$ and $\eta\approx200$, our \texttt{DUSTY} modeling suggest a mass loss rate $\dot{M}\sim10^{-5}$\,M$_{\odot}$yr$^{-1}$. This is substantially higher than what is expected for an isolated early-AGB star, e.g., from the prescription of \citep{Reimers1975}, which suggests an expected $\dot{M}\sim10^{-7}$\,M$_{\odot}$\,yr$^{-1}$. This indicates that the primary star was in a phase of enhanced mass loss before the precursor brightening began. 

Enhanced AGB mass loss in the presence of a binary companion has been studied by several studies. Primary stars close to or past Roche lobe overflow lose mass at rates $10^{-4}-10^{-3}$\,M$_{\odot}$\,yr$^{-1}$ (e.g., \cite{Ritter1988,BonavicMarinovic2007, MacLeod2020}) --- over two magnitudes larger than standard AGB winds. Even before Roche lobe overflow, tidal interactions between the AGB star and the companion can increase the mass loss rate \cite{Tout1988}. Binary systems with high eccentricity can undergo enhanced mass loss at periastron \cite{Soker2000, BonavicMarinovic2007}. Such eccentric binaries can also undergo grazing envelope evolution, in which the companion can launch jets due to enhanced accretion during periastron \citep{Soker2015, Kashi2018}. Another possibility is the wind Roche Lobe overflow scenario \cite{Mohamed2007, deVal-Borro2009,
Chen2017, Saladino2019}, where winds from the AGB star can be equatorially focused onto the secondary, resulting in accretion rates $\sim100$ times larger than standard wind mass accretion. While a fraction of this mass is accreted onto the companion, the remainder can escape from the L2 Lagrangian point and produce equatorial, spiral outflows from the system \cite{Mohamed2012}, consistent with the geometry of the CSM inferred from our spectroscopic observations (Section \ref{sec:modeling_lrn_spec}). Mass loss rates from unstable mass transfer in binary systems has been calculated by \cite{MacLeod2020, MacLeod2022}. These outflows carry out angular momentum from the binary, shrinking the orbit and bringing the stars closer to the onset of CEE. 

Our \emph{JWST} observations probe the outer layers of dust formed during this enhanced mass loss phase. The inner radius of the dust shell in the \emph{JWST} observations is $r_{dust, in}\approx3\times10^{15}$\,cm. This dust could not have condensed in material ejected in the LRN eruption, as this would imply ejecta speeds exceeding 6000\,km\,s$^{-1}$, which are not observed. This dust could also not have condensed during the precursor eruption phase, as that would require precursor ejecta velocities $\geq160$\, km \, s$^{-1}$, substantially higher than the escape velocity at the surface of the progenitor ($\approx87$\, km \, s $^{-1}$). This indicates that the dust existed even before the precursor, during the enhanced mass loss phase of the binary system. Comparing the dust optical depths and dust shell inner radii inferred from the \emph{JWST} observations to the progenitor further shows that both dust components are likely part of the same dust outflow. Assuming $\rho\propto r^{-2}$, the dust optical depth $\tau$ is inversely proportional to $r_{\rm{dust, in}}^{-1}$. Our inferred values show that $\tau_{\rm{progenitor}}$/$\tau_{JWST}=18_{-6}^{+18}$ is broadly consistent with $r_{JWST}$/$r_{\rm{progenitor}} = 12_{-5}^{+7}$. We also find that the \emph{JWST} dust lies just beyond the distance out to which pre-existing dust could be destroyed by the 2025 LRN eruption. The maximum distance within which dust can be destroyed by this eruption is given by (e.g., \cite{Metzger2023})
\begin{equation}
\begin{split}
r_{\rm{max}}=1.06\times10^{15}~\rm{cm}~\left(\frac{L_{peak}}{6\times10^{39}\,erg\,s^{-1}} \right)^{1/2}\left(\frac{Q_{r}}{0.1}\right)^{1/2} \left(\frac{a}{0.1\mu m}\right)^{-1/2} (T_{s,1700K})^{-5/2}
\end{split}
\end{equation}
where L$_{\rm{peak}}$ is the peak luminosity of the eruption, Q$_{r}$ is the dust scattering efficiency in the \emph{r} band (taken from \cite{Draine84}), $a$ is the grain size, and T$_{s}$ is the dust sublimation temperature. This suggests that the AT\,2025abao eruption only destroyed dust out to $\approx10^{15}$\,cm, lower than the dust shell inner radius derived from the \emph{JWST} observations ($\approx3\times10^{15}$\,cm). We therefore conclude that the dust probed by the \emph{JWST} observations are remnants of the pre-existing dust from enhanced mass loss in the progenitor binary, that was not sublimated by the 2025 eruption. 

The dust mass surrounding the progenitor binary before merger can be used to constrain its mass ratio. Our inferred \texttt{DUSTY} parameters can be converted to the total dust mass using (e.g., \cite{Lau2025})

\begin{equation}
\label{eqn:mdust}
M_{\rm{dust}} = \frac{4 \pi \tau_V R_\mathrm{in}^2 Y }{\kappa_V}, 
\end{equation}

The most uncertain parameter in Equation \ref{eqn:mdust} is the shell thickness Y, as we cannot determine it uniquely for the progenitor or the JWST SED. As both the progenitor and JWST SEDs probe the same dust shell, $Y_{\rm{prog}} = (r_{in, JWST}/r_{in, prog}) Y_{JWST}$. Using the values inferred from our SED modeling, we derive a dust mass around the progenitor of $M_{\rm{dust, prog}} = 6_{-4}^{+5}\times10^{-5}$\,M$_{\odot}$. 

\subsubsection{Comparisons to theoretical predictions of enhanced mass loss}
\label{sec:disc_mass_loss_comparison}
We now compare this dust mass estimate to theoretical expectations from Roche Lobe overflow from a binary system. First, we use the \texttt{python} package \texttt{RLOF} \cite{MacLeod2020} that models mass loss from a binary system prior to the onset of common-envelope evolution using a semi-analytical formalism developed by \cite{MacLeod2018b, MacLeod2020}. We make several assumptions about the progenitor binary system -- we assume a primary mass of 7\,M$_{\odot}$ and the primary radius of 350\,R$_{\odot}$, a structure coefficient of 4/3 and an adiabatic structure coefficient of 5/3, and no co-rotation of the binary system. Varying these parameters results in $\approx50\%$ variations in the mass loss rates \cite{MacLeod2020,De2023Nature}. We compute the cumulative mass lost by integrating the mass loss rates output by \texttt{RLOF} for several mass-ratios ($q = M_{\rm{secondary}}/M_{\rm{primary}}$) ranging from 0.01 to 1, and convert them to dust masses assuming a dust-to-gas ratio $X_{dg}=5\times10^{-3}$. 

Second, we computed stellar and binary models with \textsc{MESA}, version \texttt{r12115}, using a numerical setup closely following \cite{Temmink2023}, with modifications specific to the present system as described below. We adopted a Solar metallicity of $Z=0.0142$ \citep{Asplund2009}. Convection was treated using the Ledoux criterion \citep{Ledoux1947}, with a mixing-length parameter $\alpha_{\rm MLT}=1.82$, semiconvection with $\alpha_{\rm semi}=0.1$ \citep{Langer1983}, and the same overshooting prescription used in the \textsc{MIST} calibration \citep{Choi2016}. Wind mass loss was included through a Reimers prescription on the RGB with $\eta_{\rm R}=0.1$ \citep{Reimers1975} and a Bl\"ocker prescription on the AGB with $\eta_{\rm B}=0.2$ \citep{Blocker1995}. We employed an extended nuclear network, constructed from \texttt{pp\_cno\_extras\_o18\_ne22.net} with \texttt{add\_co\_burn}, in order to follow consistently the H-, He-, and C/O-burning channels relevant for these donor models. The numerical mesh was refined relative to the default settings near convective boundaries and strong composition gradients.
	
Motivated by the inferred progenitor properties, we focused on a $7\,M_\odot$ donor sequence, which provides the closest match to the allowed progenitor region in the Hertzsprung--Russell diagram. Within this sequence, we selected donor models that filled their Roche lobe at \(R_{\rm d}\simeq 340\,R_\odot\), corresponding to the observationally allowed progenitor region. For each such donor model, we considered accretor-to-donor mass ratios \(M_{\rm a}/M_{\rm d}=\{0.01,\,0.1,\,0.2,\,0.5\}\), and thus different companion masses, setting the initial orbital separation in each case with the Eggleton approximation for the equivalent-volume Roche-lobe radius \citep{Eggleton1983}. In all calculations, the initial orbit was assumed to be circular. Mass-transfer rates were computed with the \texttt{Kolb} prescription \citep{Kolb1990}. To bracket the uncertain role of mass and angular-momentum loss, we considered two limiting cases: fully conservative mass transfer, and a fully non-conservative isotropic-re-emission limit, in which all transferred material that is not retained is removed from the vicinity of the accretor with the accretor's specific orbital angular momentum \citep[e.g.][]{Soberman1997,Tauris1999}. In all calculations, the accretor was treated as a point mass.

We do not include tides or donor rotation in our fiducial calculations. The reason is that, in the regime of interest here, the decisive phase of the interaction is the rapid mass-transfer runaway after the onset of Roche-lobe overflow. Once mass transfer begins, the transfer rate rises on a timescale of only decades, such that the resulting orbital evolution is controlled primarily by the unstable mass-transfer feedback rather than by the much slower tidal redistribution of angular momentum. In this sense, the omission of tides and rotation is a deliberate simplification aimed at isolating the dominant driver of the final approach to outer-lobe overflow.

As a consistency check, we carried out a small set of additional \textsc{MESA} calculations including donor rotation and convective-envelope tides, using the \texttt{Hut\_conv} prescription, which is based on \citep{Hurley2002}. These exploratory models show that, although many systems formally enter the Darwin-unstable regime before contact, the donor is already strongly non-synchronous by that stage, indicating that tidal torques are too weak to maintain corotation during the rapid late-AGB expansion. As a result, the subsequent evolution from first Roche-lobe overflow to outer-lobe overflow is nearly unchanged relative to the non-rotating models. This supports our interpretation that, for the systems considered here, crossing the Darwin boundary signals the loss of a stable synchronized equilibrium, but does not by itself introduce a new fast inspiral timescale \citep[e.g.][]{Nandez2014}.

In addition to ordinary Roche-lobe overflow, we tracked the donor's outer-lobe radius using the fit of \cite{Temmink2023}, and terminated the one-dimensional binary calculations when the donor filled its outer lobe. Beyond the onset of outer-lobe overflow, the mass loss is expected to become strongly non-spherical and dynamically driven, such that the subsequent evolution is no longer robustly represented by hydrostatic one-dimensional calculations; this is the same motivation adopted by \cite{Temmink2023} in delimiting the regime of validity of the \textsc{MESA} binary models. At that point we stored the full stellar structure and used it to evaluate the binding energy of the remaining envelope. For a chosen mass coordinate $m$, we calculated
\begin{equation}
    E_{\rm bind}(>m)
    =
    \int_m^{M_\star}
    \left(
    -\frac{Gm_r}{r}+u
    \right)\,{\rm d}m,
\end{equation}
where $u$ is the specific internal energy. 

We compare the theoretically computed dust masses $\approx10$ years before the orbital separation becomes equal to the primary radius to our inferred progenitor dust mass from above in Extended Data Figure \ref{fig:progenitor_mdust_comparisons}. Our measured dust mass suggests that the progenitor binary has a low mass ratio q$=0.10^{+0.10}_{-0.05}$, suggesting that the companion is a low mass $\approx0.7$\,M$_{\odot}$ star. 

\subsubsection{The precursor : A slow inspiral followed by a rapid dynamical plunge}
\label{sec:disc_precursor}
Following the quiescent phase, the system started brightening slowly, starting in 2018. This precursor phase differs phenomenologically from other LRNe where it has been well-studied. First, the optical lightcurves of other LRNe such as V1309Sco, M31-LRN-2015, and AT\,2019zhd show that they brightened by several magnitudes over their quiescent luminosities during their precursor phase, while AT\,2025abao only brightened by $\approx1$\,mag (Extended Data Figure \ref{fig:lc_precursor_comparisons}). Second, the AT\,2025abao precursor phase lasted about seven years, corresponding to $<10$ orbital periods of the system (P$_{\rm{orb}}\approx280$\,days for a low mass companion orbiting close to the surface of the primary, and larger for companions further away). This is substantially shorter than in other well-studied LRN precursors such as V1309Sco and M31-LRN-2015, where the slow precursor brightening lasted for over a hundred orbits (see Figure \ref{fig:precursor_comparisons}). 

The precursor brightening seen in LRNe has generally been attributed to outflows from the binary system, e.g., from L2 mass loss, that can produce a slow brightening of the binary over several hundred orbits \citep{Pejcha14, Pejcha2017ApJ, Blagorodnova2020}. However, these outflows cannot power the precursor brightening in AT\,2025abao. The kinetic luminosity of the outflows is given by 
\begin{equation}
    L_{kin} = \frac{1}{2} \dot{M} v_{\rm{outflow}}^{2} \approx 10^{36} \rm{erg}\,\rm{s}^{-1} \left(\frac{v_{\rm{outflow}}}{0.2 v_{\rm{esc}}}\right)^{2} \left(\frac{\dot{M}}{10^{-2}\,M_{\odot}\,yr^{-1}}\right)
    \label{eqn:lkin}
\end{equation}
where the radiative efficiency $\eta\ll1$ \cite{Pejcha14}. This is significantly lower than the precursor luminosity increase of $\gtrapprox10^{38}$\,erg\,s$^{-1}$ observed for AT\,2025abao, even for mass loss rates as high as 1\,M$_{\odot}$\,yr$^{-1}$. Furthermore, this is also lower than the quiescent progenitor luminosity ($\approx7\times10^{37}$\,erg\,s$^{-1}$), showing that outflows are not energetic enough to alter the progenitor's brightness. As noted by \cite{Pejcha14}, slow brightening due to L2 mass loss, like that seen in V1309Sco, is only expected for short-period binaries (P$\leq30$\,days), where the outflow kinetic energy is high relative to the binary luminosity. This condition is true for LRNe with main-sequence or HG progenitors such as V1309Sco and M31-LRN-2015, and thus L2 mass loss is likely responsible for slow brightening by several magnitudes over $\sim$hundreds of orbital timescales that precede their main eruption. In contrast, for binaries like AT\,2025abao, L2 outflows pile up around the binary but are not energetic enough to increase the binary luminosity. This is illustrated in Figure \ref{fig:precursor_comparisons}, which shows the precursor luminosities and durations as functions of the escape velocity from the primary star. AT\,2025abao has the lowest primary escape velocity and the shortest precursor duration (in orbital period units) and cannot be powered by L2 mass loss (gray dashed lines). 

Instead, the AT\,2025abao precursor is consistent with being powered by energy dissipation due to frictional drag on the companion, as it spirals through the outer envelope of the primary star. The rate of this energy dissipation has been calculated in the context of planetary engulfment by giant stars (e.g., \cite{Ostriker1999, Metzger2012, MacLeod2015, MacLeod2018, SoumiDe2020, OConnor2023}), and can be calculated as 

\begin{equation}
    L_{\rm{decay}} = F_{\rm{drag}}.v_{\rm{orb}} = C_{d} \pi R_{a}^2 \rho v_{\rm{orb}}^{3}
\end{equation}
where $C_{d}$ is a drag coefficient of order unity, $\rho$ is the envelope density at the radius at which the companion is inspiraling, $v_{\rm{orb}}$ is the corresponding orbital speed, and $R_{a}$ is the gravitationally enhanced cross-sectional radius given by $R_{a} = \frac{2GM_2}{v_{\rm{orb}}^2}$ where $M_{2}$ is the companion mass. This expression therefore simplifies to 
\begin{equation}
    L_{\rm{decay}} = 4\pi G^{3/2} C_{d} M_{2}^{2} \rho a^{1/2} (M_1 + M_2)^{-1/2}
    \label{eqn:ldecay}
\end{equation}
where M$_{1}$ is the mass of the primary and $a$ is the orbital radius where the companion dissipates energy. Assuming $\rho(r) \propto r^{-3}$ in the outer envelope, and the density at the photospheric radius $R_1$ is $\rho_1 \sim 0.1 \frac{3M_1}{4\pi R_1^3}$ \cite{MacLeod2018}, the energy dissipation rate is given by 
\begin{equation}
\begin{split}
    L_{\rm{decay}} \approx 3\times10^{38}\,\rm{erg}\,\rm{s}^{-1} \left(\frac{M_2}{0.7M_{\odot}} \right)^2 \left(\frac{\rho_1}{2\times10^{-8}{\rm g\,cm^{-3}}} \right) \left(\frac{R_1}{350 R_{\odot}}\right)^{3} \\ \left(\frac{a}{350\,\rm{R_{\odot}}} \right)^{-5/2} \left(\frac{M_1 + M_2}{7 + 0.7\,M_{\odot}}\right)^{-1/2}
\end{split}
\end{equation}
for a companion spiraling at the primary's photosphere, which is consistent with the median observed precursor luminosity. As noted by \cite{Karambelkar2025}, the characteristic timescale over which this energy is dissipated is $|E_{\rm{orb}}|/L_{\rm{decay}}\approx1000$\,days, corresponding to about five orbits of the system, consistent with the observed durations of the precursor bumps. 

Our observations show that for most of the precursor phase, the star puffed up at constant photospheric temperature to approximately 2.5 times its initial size.  
This quasi-static phase is also radiatively efficient --- as the drag luminosity $L_{\rm{decay}}$ exceeds the energy required to unbind material and launch outflows ($\sim\dot{M}v_{esc}^{2}$) --- most ($\gtrsim90\%$) of the drag luminosity is radiated. As a reference for the orbital separations involved, attributed the observed excess precursor luminosity entirely to $L_{\rm{decay}}$ suggests that the companion inspiraled from an orbital separation $\approx$1.5$R_{1}$ at the start of the precursor to $\approx$0.9$R_{1}$ at the end of the quasi-static phase. The total excess energy radiated during this quasi-static phase is $\approx 10^{46}$\,erg, consistent with the orbital energy lost during this inspiral in the primary's outer envelope. 

Identifying the precursor variations with a companion inspiral provides an independent constraint of the binary mass ratio ($q=M_2/M_1$). A lower limit comes by requiring that a companion inspiraling at the photosphere dissipates energy comparable to the progenitor luminosity, i.e. $L_{\rm{decay}}(R_1) > L_{\rm{progenitor}}$, which gives $q>0.05$. Similarly, assuming that majority of the orbital energy is radiated away, the maximum luminosity of the precursor can be used to impose an upper limit on the companion mass, as $L_{\rm{decay}} (R_1) < 2 L_{\rm{precursor}}^{\rm{max}}$, which gives $q <0.2$. These constraints agree well with those determined from the quiescent mass loss measurements above.

Our SPHEREx SED provides an estimate of the mass lost from the system during this inspiral. Our inferred dust parameters (Section \ref{sec:modeling_spherex}) and Equation \ref{eqn:mdust} correspond to a dust mass during the precursor phase M$_{\rm{dust, precursor}}=3_{-2}^{+3}\times10^{-4}$\,M$_{\odot}$. For $\eta_{gd}\approx200$, this corresponds to a total mass $M_{\rm{precursor}}=0.06^{+0.06}_{-0.04}$\,M$_{\odot}$. This estimate is broadly consistent with a CBM mass of $\sim0.1$\,M$_{\odot}$ inferred from the spectroscopic properties of the LRN eruption in Section\,\ref{sec:disc_lrn_eruption}. Theoretically, we estimate the mass loss rate during the precursor phase by assuming that it is proportional to the rate at which mass is gravitationally swept up by the companion \cite{Metzger2012} 
\begin{equation}
\begin{split}
    \dot{M}_{\rm{outflow}} \approx 0.14 \rm{M}_{\odot}\,\rm{yr}^{-1} \beta_{ml} \left(\frac{M_2}{0.7\,M_{\odot}} \right)^{2} \left(\frac{R_1}{350R_{\odot}}\right)^{3/2} \\ \left(\frac{\rho_1}{2.10^{-8}g/cm^3} \right) \left(\frac{a}{R_1}\right)^{-3/2} \left(\frac{M_1 + M_2}{7+0.7M_{\odot}}\right)^{-3/2}
\end{split}
\end{equation}
where we define $\beta_{ml}$ as a mass-loss efficiency. For $\beta_{ml}\sim0.1$ and a precursor phase duration of $\sim7$ years suggests $\sim0.1$\,M$_{\odot}$ expelled during this phase, broadly consistent with the observed estimates.  

These observations of a quasi-static, radiatively efficient inspiral in the primary's outer envelope closely match the expectations of theoretical CEE studies that include the effects of convection \cite{Wilson2019, Wilson2020}. These studies show that, as the companion inspirals through the outer convective envelope, excess energy can be efficiently transported to the stellar surface by convection and radiated away. The outer inspiral is therefore expected to be radiatively efficient and not contribute significantly to launching outflows and unbinding the envelope \cite{Wilson2019}. Theoretical lightcurve models of the inspiral phase in such systems predict timescales and luminosities broadly consistent with those observed in the AT\,2025abao precursor \cite{Noughani2026}.

Our observations of the final $\approx300$\,days of the precursor phase show a drop in the photospheric temperature. As $L_{\rm{decay}}\propto a^{-5/2}$, energy dissipation increases rapidly as the companion spirals further into the star and eventually the companion can no longer respond quasi-statically. This phase therefore likely corresponds to a plunge of the companion deep within the primary's envelope, eventually culminating with the LRN eruption. The duration of this phase corresponds to $\sim6$ dynamical timescales of the envelope.

These observations agree well with CEE simulations involving AGB primaries, which predict a rapid plunge-in phase lasting a few hundred days (e.g. \cite{Staff2016, Sand2020,  Glanz2021, GonzalezBolivar2022, Gonzalez-Bolivar2024, Bermudez2024}). A possibly relevant case is one where the initial binary is eccentric \cite{Staff2016, Glanz2021}. In such cases, the orbital separation oscillates on timescales of $\sim$hundreds of days in addition to decreasing rapidly (e.g., Fig.\,7 in \cite{Staff2016}) as the binary circularizes while undergoing CEE. It is reasonable to expect this behavior to be imprinted on the precursor luminosity. As noted by \cite{Karambelkar2025}, the \emph{r}-band precursor of AT\,2025abao shows modulations on three distinct timescales -- a long bump lasting 1500 days, followed by two small-amplitude bumps lasting $\approx84$ days each, followed by the second long-duration bump of $\approx1000$ days. These timescales could be indicative of an eccentric orbit for the progenitor binary. Future detailed theoretical modeling of the precursor phase of AT\,2025abao will help explore this possibility. 

In summary, our modeling of the precursor phase shows that it was powered by a low mass companion spiraling through the outer envelope that eventually underwent a dynamical plunge-in deep into the stellar envelope. 

\subsubsection{The LRN eruption: Termination of the rapid plunge}
\label{sec:disc_lrn_eruption}
Following the precursor phase, the source rapidly brightened and transitioned to the LRN eruption with fast moving ejecta, marking the culmination of CEE. The rise time to peak luminosity is governed by the photon diffusion timescale through this fast ejecta, and is given by (e.g., \cite{Arnett80, Padmanabhan2001, MacLeod17, Matsumoto2022})
\begin{equation}
    t_{\rm{pk}} \approx 6.7 \rm{d}~\left(\frac{M_{ej}}{10^{-2}M_{\odot}}\right)^{\frac{1}{2}} \left(\frac{v_{\rm{fast}}}{500 \rm{km s}^{-1}}\right)^{-\frac{1}{2}}
\end{equation}

For $t_{pk}\approx20-25$\,days and $v_{\rm{fast}}\approx800-900$\,km\,s$^{-1}$ from our early time spectra, we get $M_{ej}\approx0.14-0.25$\,M$_{\odot}$. The expected peak luminosity from cooling of this expanding ejecta is $\approx2\times10^{39}$\,erg\,s$^{-1}$ \cite{Matsumoto2022}, consistent with the observed peak luminosity at the end of the initial rise. As outlined by \cite{Metzger2017, MacLeod17}, this initally ejected material is likely spherically symmetric and expands freely in the polar direction, but can be obstructed by pre-existing circumbinary material (CBM) in the equatorial plane .

Our spectra show clear signs of shock interactions of the fast-moving ejecta with CBM. The hydrogen emission line velocity decreases from $\approx900$\,km\,s$^{-1}$ to $\approx150$\,km\,s$^{-1}$. This velocity deceleration matches the expectation for momentum-conserving collisions between fast-moving ejecta and CBM with a density profile $\rho \propto r^{-2}$ (Figure \ref{fig:opt_spec}) \cite{Metzger2017}. The total CBM swept up is about five times the initial ejecta mass. Assuming that the CBM is distributed in a disk that covers 20\% of the solid angle subtended by the initial spherically symmetric ejecta, the total mass of CBM swept up is $\approx0.1-0.2$\,M$_{\odot}$, comparable with the mass of the fast-moving ejecta. These estimates are also consistent with the mass lost during the precursor phase estimated in Section \ref{sec:disc_precursor}. The total kinetic energy dissipated is $\sim3\times10^{47}$\,erg, indicating that these shock-interactions are energetic enough to power the observed plateau that radiated a total energy $\approx4\times10^{46}$\,erg. This confirms theoretical studies of the role of shock-interactions in powering LRN lightcurves \cite{Metzger2017, Kirilov2025}.

\subsubsection{Constraining the outcome of CEE in AT 2025abao}
\label{sec:disc_merger}
The AGB progenitor primary of AT\,2025abao differs from those of other LRNe, that have main sequence (MS) or Hertzsprung Gap (HG) primaries \cite{Pastorello2019a, Blagorodnova2021}. MS or HG stars have tightly bound radiative envelopes, and companions that are engulfed in these envelopes are expected to merge with the primary star's core \cite{Klencki2021, MacLeod2022}, while the loosely bound convective envelope of the AGB progenitor primary of AT\,2025abao makes a successful envelope ejection possible. However, several pieces of evidence indicate that CEE in AT\,2025abao culminated in a merger. 

First, the kinetic energy of the fast moving ejecta are $\sim10^{48}$\,erg. Assuming orbital energy decay is the major energy reservoir, this requires the low mass (q$\sim0.1$) companion to have inspiraled down to a radius of $\approx$3R$_{\odot}$ (Figure \ref{fig:precursor_comparisons}), which coincides with the radius of the helium core of the primary. Second, the total unbound mass during the precursor and eruption phases is $\sim0.2-0.4$\,M$_{\odot}$, which is a small fraction ($\approx10\%$) of the primary envelope, suggesting that most of the primary envelope has remained intact after the event. Third, the main eruption of AT\,2025abao is remarkably similar to other LRNe (Extended Data Figure \ref{fig:lrn_properties_comparison}), which are known to result from stellar mergers. Finally, the transition from the slow variations during the inspiral to the rapid brightening of the main eruption to high luminosities ($>10^{39}$\,erg\,s$^{-1}$) points to a sudden energy deposition episode within the common envelope, which would be expected in case of a violent merger of the companion with the primary core. 

While this evidence points towards a stellar merger, we discuss future observations that would conclusively rule out a tight surviving binary that is gradually inspiraling within and stripping away the remnant envelope through a self-regulated inspiral, as hypothesized by several theoretical studies \cite{Meyer1979, Podsiadlowski2001, Clayton2017, Gagnier2023, Tuna_2023}. First, sustained mass loss from a surviving binary would likely result in a slow or no evolution of the dust shell radius inferred from late-time IR observations \cite{Karambelkar2026_jwst}. In contrast, a merger would have a shell expanding at speeds comparable to those inferred from the eruption, as has been seen in other LRNe \cite{Karambelkar2026_jwst}. Second, using Equation\,\ref{eqn:ldecay}, the energy deposition rate of the companion inspiraling at an orbital separation of $\approx10$\,R$_{\odot}$ is $L_{\rm{decay}}\approx10^{37}$\,erg\,s$^{-1}$. Since this luminosity is comparable to that of the progenitor primary star — and LRN remnants remain overluminous relative to their progenitors even a decade after eruption \cite{Karambelkar2026_jwst} — observations conducted decades later could be sensitive to this energy dissipation. Third, additional energy deposition from a surviving binary would also provide mechanical power to the ejecta, affecting their kinematics. Extended Data Figure \ref{fig:velocity_evolution_additional_energy} shows the expected spectral velocity evolution, similar to Section \ref{sec:modeling_lrn_spec}, but by including an additional mechanical power source of magnitude $\dot{E}$. The observed velocity evolution from our spectra constrains the mechanical power deposited by such a binary to $\dot{E}<10^{40}$\,erg\,s$^{-1}$. For context, the orbital energy of a binary comprising the primary core and the companion at a separation of $\approx10$\,R$_{\odot}$ is 
$\approx-10^{47}$\,erg. Assuming all dissipated orbital energy is converted to mechanical power, this limit implies that the orbital decay timescale of a surviving binary exceeds $\approx$100 days. Spectral line profile monitoring over a longer temporal baseline will probe lower values of $\dot{E}$, corresponding to longer decay timescales. Finally, dynamical instabilities can produce large amplitude pulsations in the surviving envelope \cite{Clayton2017}, which should be detectable by future photometric monitoring of AT\,2025abao.

\subsubsection{The energetics of common-envelope evolution}
\label{sec:disc_alpha}
Assuming that orbital energy decay constitutes the main energy reservoir, the energy budget of the CEE event in AT\,2025abao can be expressed as 
\begin{equation}
    \Delta E_{\rm{orb}} = E_{\rm{unbind}} + E_{\rm{rad}} + KE_{\rm{ej}}
\end{equation}
where $\Delta E_{\rm{orb}}$ is the total orbital energy lost by the inspiraling companion, $E_{\rm{unbind}}$ is the energy expended in unbinding part of the envelope, $E_{\rm{rad}}$ is the energy radiated away, and $KE_{\rm{ej}}$ is the kinetic energy imparted to the ejecta. We plot these quantities inferred for AT\,2025abao from our observations in Figure \ref{fig:precursor_comparisons}. We find that during the rapid plunge, almost all $\Delta E_{\rm{orb}}$ ($\approx10^{48}$\,erg) was transferred to the kinetic energy of a small amount of unbound material that requires $E_{\rm{unbind}}\approx10^{46}$\,erg to unbind them. During this phase, only a small fraction of $\Delta E_{\rm{orb}}$ was radiated away ($E_{\rm{rad}}\approx10^{46}$\,erg). We note that this differs from the quasi-static slow inspiral phase, during which most of the energy dissipated due to frictional drag was radiated away (Section \ref{sec:disc_precursor}). Therefore, it appears that the initially radiatively efficient slow inspiral that powered the slow brightness variations rapidly transitioned to a kinetically efficient plunge that powered the rapid eruption.

We now examine AT\,2025abao in the context of the energy formalism of CEE \cite{vandenHeuvel1976,Tutukov1979,Webbink84, Ivanova2013araa} that is widely used in binary population synthesis models to predict the outcome of CEE in binary systems. In this formalism, a binary is considered to have survived the CEE by successfully ejecting the common envelope if
\begin{equation}
    E_{\rm{bind}}(a_{\rm{f}}) < \alpha \Delta E_{\rm{orb}}(a_{\rm{f}})
    \label{eqn:energy_formalism}
\end{equation}
holds true for a value of $a_{f}$, the final orbital separation, exceeding the core radius of the primary. $E_{\rm{bind}}$ is the binding energy of the envelope exterior to $a_{f}$, $\Delta E_{\rm{orb}}(a_{\rm{f}})$ is the orbital energy lost in an inspiral to $a_{\rm{f}}$, and $\alpha$ is an efficiency parameter representing the fraction of orbital energy that goes into unbinding envelope. Observational constraints on $\alpha$ have primarily been derived by backward modeling of post-CE binary systems (e.g., \cite{Nelemans2000, Nelemans2005,Maxted2006, Afsar2008, Webbink2008, Zorotovic2010, deMarco2011, Davis2012, Scherbak2023}. Although this method is limited by several uncertainties about binary evolution \cite{deMarco2011}, the indicative constraints obtained by modeling close double white dwarf (WD) systems and WD-M dwarf binaries suggest low values of $\alpha$ in the range $\sim0.1-0.3$ \cite{Zorotovic2010, Scherbak2023}.  

As seen in Figure \ref{fig:precursor_comparisons}, the dissipated orbital energy exceeds the envelope binding energy for orbital separations lower than $20$\,R$_{\odot}$. Adopting values of $\alpha\approx1$ would therefore suggest that the binary survived at this orbital separation, inconsistent with our observations that point to an even deeper inspiral (to $\approx3$\,R$_{\odot}$) resulting in a stellar merger. Building on this, we use the progenitor binary parameters inferred for AT\,2025abao to constrain the values of $\alpha$ that would make Equation \ref{eqn:energy_formalism} consistent with the observed merger outcome of its CEE. 

We determine the binding energy of the primary envelope as a function of radius from our MESA model (as defined in Section \ref{sec:disc_mass_loss_comparison} and plotted in Figure \ref{fig:precursor_comparisons}). We note that our binding energy calculation includes the envelope's internal energy of the envelope (as in \cite{Scherbak2023}). We compare this with the dissipated orbital energy of a low mass $q=0.1_{-0.05}^{+0.10}$ through this envelope. Our observations indicate the companion inspiraled to merge with the helium core (R$\approx3$\,R$_{\odot}$), resulting in a dissipated orbital energy $\approx8\times10^{47}$\,erg. The binding energy of the envelope outside this radius is $\approx1.6\times10^{47}$\,erg. A stellar merger implies Equation \ref{eqn:energy_formalism} is never true for $a_{\rm{f}}>3$\,R$_{\odot}$, which implies $\alpha < 0.2_{-0.1}^{+0.2}$. The uncertainty in this upper limit comes from the uncertainty on the companion mass (Figure \ref{fig:precursor_comparisons}). Our upper limit of $\alpha \lesssim 0.4$ is broadly consistent with the values inferred from the post-CE binary systems described above. In a similar vein, an absolute lower limit on $\alpha$ can be inferred from our observations using the inferred ejecta mass. Our observations show that a $\sim0.2-0.4$\,M$_{\odot}$ was unbound from the system. The energy required to unbind the outermost $\sim0.2-0.4$\,M$_{\odot}$ of the envelope is $\approx10^{46}$\,erg. Comparing this to the dissipated orbital energy shows that $\alpha\gtrsim0.01$. 

We compare this direct constraint on $\alpha$ with that derived from modeling post-CE binary systems. We use the mass-ratios and $\alpha$ values inferred for the progenitors of such systems by \cite{deMarco2011} in Figure \ref{fig:precursor_comparisons}. As the $\alpha$ values reported by \cite{deMarco2011} do not include the contribution of internal energy, we multiply their reported values by 0.5 to include its dependence on the envelope binding energy (Section 4 of \cite{deMarco2011}). While the uncertainties on the post-CE $\alpha$ measurements are high, the values are broadly consistent with our constraint. 

The low values of $\alpha$ inferred for post-CE systems ($\sim$0.01-0.3 \cite{Zorotovic2010, deMarco2011}) are consistent with a picture in which a substantial fraction of the released orbital energy is ultimately not available to unbind the envelope. Such an effective reduction in the coupling of orbital energy to envelope ejection could arise for several reasons, including energy being transported outward and radiated, energy going into other forms of envelope motion or expansion, or uncertainties in the envelope binding energy used in the standard energy formalism. Efficient convection in the extended AGB envelope provides one physically motivated mechanism for transporting deposited orbital energy to the surface, where it can be radiated away \cite{Wilson2019, Wilson2020, Noughani2026}. The observed transition from a prolonged radiatively efficient phase to a rapid plunge provides additional evidence consistent with this later interpretation.

We note that while the energy formalism is a useful parametric prescription for predicting the outcomes of CEE, our observations demonstrate that it fails to fully capture the underlying physics of CEE. The $\alpha$-prescription assumes that the transferred orbital energy is primarily expended towards unbinding the envelope and ejecting it at the escape velocity of the primary star. However, our observations show that only $\approx1\%$ of the orbital energy was used to unbind a small fraction of the envelope, and a large fraction was expended to accelerate these ejecta to speeds much higher than the primary escape speeds. Future detailed simulations will shed light on the physics of CEE in AT\,2025abao. 

The progenitor binary system of AT\,2025abao resembles the progenitors of post-CE binaries hosting massive white dwarfs \cite{Yamaguchi2024b, Landsman1993, Belloni2024}. \cite{Motherway2026} modeled 2M07515777+1807352 -- a binary system comprising a 1.08\,M$_{\odot}$ WD with a 0.66\,M$_{\odot}$ MS companion -- and found that it likely formed through a CEE episode when the massive primary ($\approx6$-$7$\,M$_{\odot}$) was in its early or thermally pulsing AGB phase. The inferred progenitor masses for this system are thus remarkably similar to those of AT\,2025abao. If confirmed by future observations, our identification of a likely merger in AT\,2025abao would suggest that the formation of such systems requires CEE to occur during the thermally pulsing AGB phase, when the envelope is even more extended than that of the early-AGB primary in AT\,2025abao. 

In summary, our observations of AT\,2025abao present an end-to-end characterization of a system undergoing CEE. Extended Data Figure \ref{fig:graphic} shows a graphical representation of the different phases of AT\,2025abao presented. Looking ahead, searches for CE transients that leverage synergies between optical and infrared time-domain surveys, such as the Vera Rubin Observatory \cite{Ivezic2019} and the Nancy Grace Roman Space Telescope are poised to uncover the diversity of CEE outcomes. These surveys have the sensitivity to characterize CE transients similar to AT\,2025abao in $\approx100$ Milky Way-like galaxies to a distance of $\approx15$\,Mpc \cite{Noughani2026, Karambelkar2023, Howitt2020}.

\newpage
\section*{Data Availability}
All ground-based optical and near-infrared photometry and spectra, together with the JWST and SPHEREx SEDs presented in this paper, are available for download from Zenodo (doi: 10.5281/zenodo.22948026). This repository also contains the bolometric parameters inferred from our SED fits. The JWST images are publicly available through MAST (doi: 10.17909/07j6-dq13), and the \emph{SPHEREx} Quick Release 2 (QR2) images are publicly available through IRSA. 

\section*{Code Availability}
Code used for the analysis of spectra and SEDs used in this paper is available on Github at \url{https://github.com/virajkaram/at2025abao/}. V.K. will provide any additional code or data, upon request.

\section*{Acknowledgements}
We thank Jim Fuller, Wynn Jacobson-Galan, Zhuo Chen, David Hogg, and Julianne Dalcanton for useful discussions. V.K. was supported by NASA through the NASA Hubble Fellowship grant \#HST-HF2-51578.001-A awarded by the Space Telescope Science Institute, which is operated by the Association of Universities for Research in Astronomy, Inc., for NASA, under contract NAS5-26555. V.K. also acknowledges the hospitality of the CCA-Flatiron Institute. B.D.M. acknowledges support from NASA (grant 80NSSC24K0934, 80NSSC26K0300) and the National Science Foundation (grant AST-2406637).  The Flatiron Institute is supported by the Simons Foundation. N.B., M.A.GM, G.G.M., H.T., M.W., and G.K. acknowledge being funded by the European Union (ERC, CET-3PO, 101042610). Views and opinions expressed are, however, those of the author(s) only and do not necessarily reflect those of the European Union or the European Research Council Executive Agency. Neither the European Union nor the granting authority can be held responsible for them. The publication is part of the research project PID2024-155585NA-I00 funded by MICIU/AEI /10.13039/501100011033. N.B., M.A.GM, G.G.M., H.T., M.W., and G.K. also acknowledge financial support from grant CEX2024-001451-M funded by MICIU/AEI/10.13039/501100011033. D.F.’s contribution to this material is based upon
work supported by the National Science Foundation un-
der Award No. AST-2401779. G.C.A. thanks the Indian National Science Academy (INSA) for support under the INSA Senior Scientist programme. C.L. is supported by DoE award \#\,DE-SC0025599. SdW is supported by a research grant (VIL60862) from VILLUM FONDEN. D.J. acknowledges support from the Agencia Estatal de Investigaci\'on del Ministerio de Ciencia, Innovaci\'on y Universidades (MCIU/AEI) and the European Regional Development Fund (ERDF) with reference PID-2022-136653NA-I00 (DOI:10.13039/501100011033). D.J. also acknowledges support from MCIU/AEI and the European Union NextGenerationEU/PRTR with reference CNS2023-143910 (DOI:10.13039/501100011033). 

This publication makes use of data products from the Wide-field Infrared Survey Explorer, which is a joint project of the University of California, Los Angeles, and the Jet Propulsion Laboratory/California Institute of Technology, funded
by the National Aeronautics and Space Administration. This publication makes use of data products from the Spectro-Photometer for the History of the Universe, Epoch of Reionization and Ices Explorer (SPHEREx), which is a joint project of the Jet Propulsion Laboratory and the California Institute of Technology, and is funded by the National Aeronautics and Space Administration. We acknowledge the support of the National Aeronautics and Space Administration through ADAP grant number 80NSSC24K0663. The computations reported in this paper were (in part) performed using resources made available by the Flatiron Institute. Astronomer observing with the Infrared Telescope Facility, which is operated by the University of Hawaii under contract 80HQTR24DA010 with the National Aeronautics and Space Administration. 

Based on observations obtained with the Samuel Oschin Telescope 48-inch and the 60-inch Telescope at the Palomar Observatory as part of the Zwicky Transient Facility project. ZTF is supported by the National Science Foundation under Grants No. AST-1440341, AST-2034437, and currently Award \#2407588. ZTF receives additional funding from the ZTF partnership. Current members include Caltech, USA; Caltech/IPAC, USA; University of Maryland, USA; University of California, Berkeley, USA; Cornell University, USA; Drexel University, USA; University of North Carolina at Chapel Hill, USA; Institute of Science and Technology, Austria; National Central University, Taiwan, German Center for Astrophysics, Germany, and OKC, University of Stockholm, Sweden. Operations are conducted by Caltech's Optical Observatory (COO), Caltech/IPAC, and the University of Washington at Seattle, USA.  The ZTF forced-photometry service was funded under the Heising-Simons Foundation grant \#12540303 (PI: Graham). The Gordon and Betty Moore Foundation, through both the Data-Driven Investigator Program and a dedicated grant, provided critical funding for SkyPortal. SED Machine is based upon work supported by the National Science Foundation under Grant No. 1106171. 

WINTER’s construction is made possible by the National Science Foundation under MRI grant number AST-1828470 with early operations supported by AST-2206730. Significant support for WINTER also comes from the California Institute of Technology, the Caltech Optical Observatories, the Bruno Rossi Fund of the MIT Kavli Institute for Astrophysics and Space Research, the David and Lucille Packard Foundation, and the MIT Department of Physics and School of Science.

Some of the data presented herein were obtained at Keck Observatory, which is a private 501(c)3 non-profit organization operated as a scientific partnership among the California Institute of Technology, the University of California, and the National Aeronautics and Space Administration. The Observatory was made possible by the generous financial support of the W. M. Keck Foundation. The authors wish to recognize and acknowledge the very significant cultural role and reverence that the summit of Maunakea has always had within the Native Hawaiian community. We are most fortunate to have the opportunity to conduct observations from this mountain.

Based on observations obtained at the international Gemini Observatory, a program of NSF NOIRLab, which is managed by the Association of Universities for Research in Astronomy (AURA) under a cooperative agreement with the U.S. National Science Foundation on behalf of the Gemini Observatory partnership: the U.S. National Science Foundation (United States), National Research Council (Canada), Agencia Nacional de Investigaci\'{o}n y Desarrollo (Chile), Ministerio de Ciencia, Tecnolog\'{i}a e Innovaci\'{o}n (Argentina), Minist\'{e}rio da Ci\^{e}ncia, Tecnologia, Inova\c{c}\~{o}es e Comunica\c{c}\~{o}es (Brazil), and Korea Astronomy and Space Science Institute (Republic of Korea).

This work is based in part on observations made with the NASA/ESA/CSA James Webb Space Telescope. The data were obtained from the Mikulski Archive for Space Telescopes at the Space Telescope Science Institute, which is operated by the Association of Universities for Research in Astronomy, Inc., under NASA contract NAS 5-03127 for JWST. These observations are associated with program \#12487. Support for program \#12487 was provided by NASA through a grant from the Space Telescope Science Institute, which is operated by the Association of Universities for Research in Astronomy, Inc., under NASA contract NAS 5-03127.

We thank the HCT-TAC for allocating time for the ToO observations. We 
thank the staff of IAO, Hanle, and CREST, Hosakote, that made these 
observations possible. The facilities at IAO and CREST are operated by 
the Indian Institute of Astrophysics, Bangalore. The GROWTH India Telescope (GIT) is a 70 cm telescope with a 0.7 degree field of view, set up by the Indian Institute of Astrophysics (IIA) and the Indian Institute of Technology Bombay (IITB) with funding from Indo-US Science and Technology Forum and the Science and Engineering  Research Board, Department of Science and Technology, Government of India. It is located at the Indian Astronomical Observatory (Hanle), operated by IIA. We acknowledge funding by the IITB alumni batch of 1994, which partially supports operations of the telescope. Telescope technical details are available at \href{https://sites.google.com/view/growthindia/}{https://sites.google.com/view/growthindia/}

\section*{Author contributions}
V.K. led the follow-up observations, carried out the analysis, and wrote the manuscript. K.D. assisted with coordination and interpretation of the follow-up observations and wrote the manuscript. K.T. created the theoretical MESA models to understand mass-loss in the progenitor binary. B.D.M. and M.M. led the theoretical interpretation of this source. N.B., G.C.A., S.B., J.B., V.B., A.B., S.C., K.D., C.F., D.F., M.A.G.M., SK.H., D.J., M.K., G.K., C.L., N.P.L., Z.M., G.M., G.G.M., N.R., R.M.R, S.R., R.A.S., H.T., M.W., S.d.W., J.W. contributed optical/near-infrared photometric and spectroscopic follow-up observations. J.J. and R.L. assisted with the analysis of the JWST observations. R.A.S., D.F., N.P.L., R.S., G.M., V.K. are builders of the WINTER observing system and contributed to survey operations during the observations presented here. S.L.G., J.C.J., R.L.R., J.P. are builders of the ZTF system and contributed to survey operations during the observations presented here. All authors contributed to the scientific interpretation.

\section*{Competing Interests}
The authors have no competing interests.

\section*{Correspondence}
Correspondence and requests for materials
should be addressed to V.K.~(email: vk2588@columbia.edu).

\subsection{Extended Data Figures}
\captionsetup[figure]{name=Extended Data Figure}
\setcounter{figure}{0}
\begin{figure}[H]
    \centering
    \includegraphics[width=\linewidth]{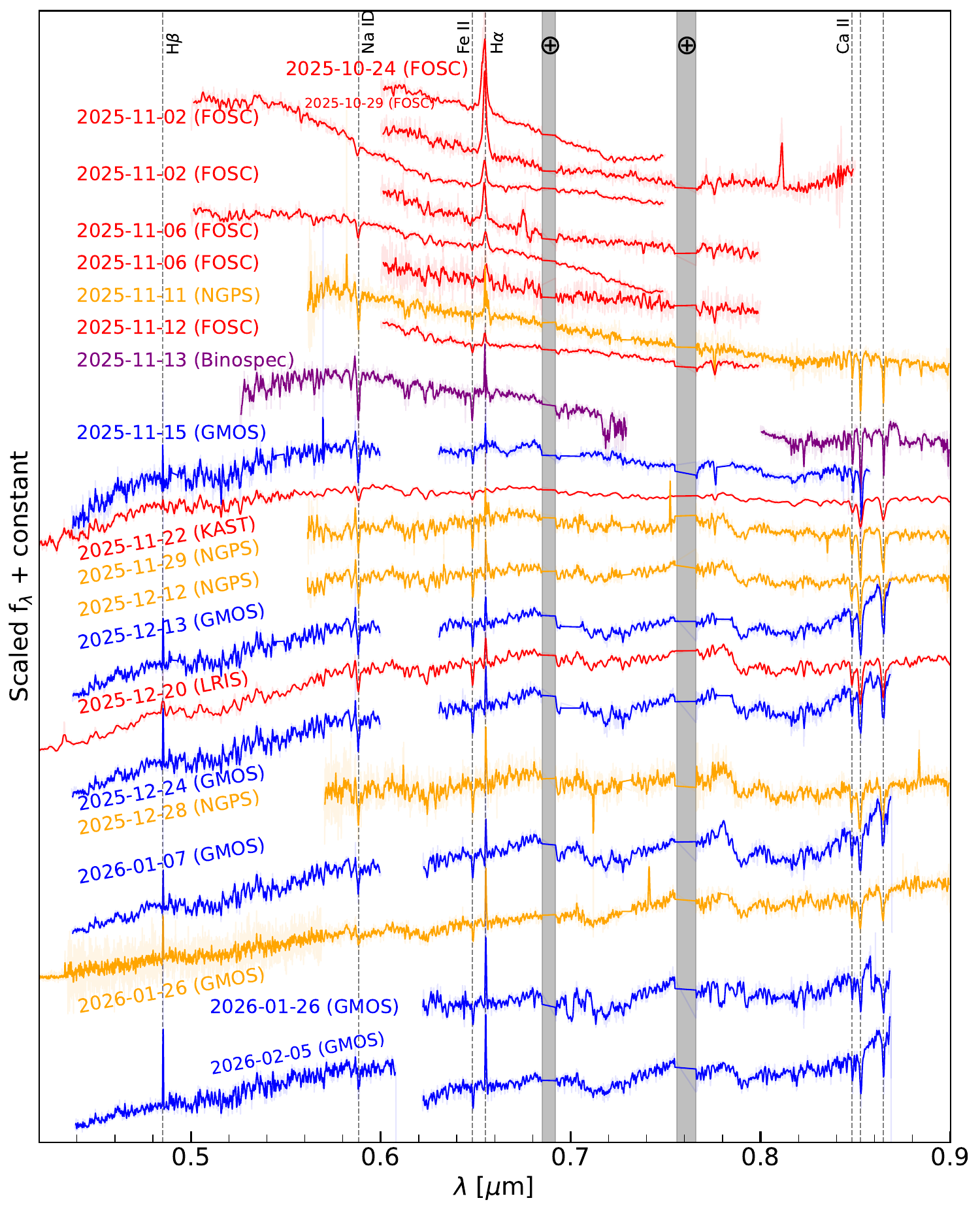}
    \caption{\emph{Top:} Optical spectra of the AT\,2025abao eruption. Its spectral evolution matches the characteristic evolution seen in LRNe --- blue continuum with emission lines at early times that transitions to a red continuum with molecular absorption features at late times.}
    \label{fig:all_optical_spectra}
\end{figure}

\begin{figure}[H]
    \centering
    \includegraphics[width=\linewidth]{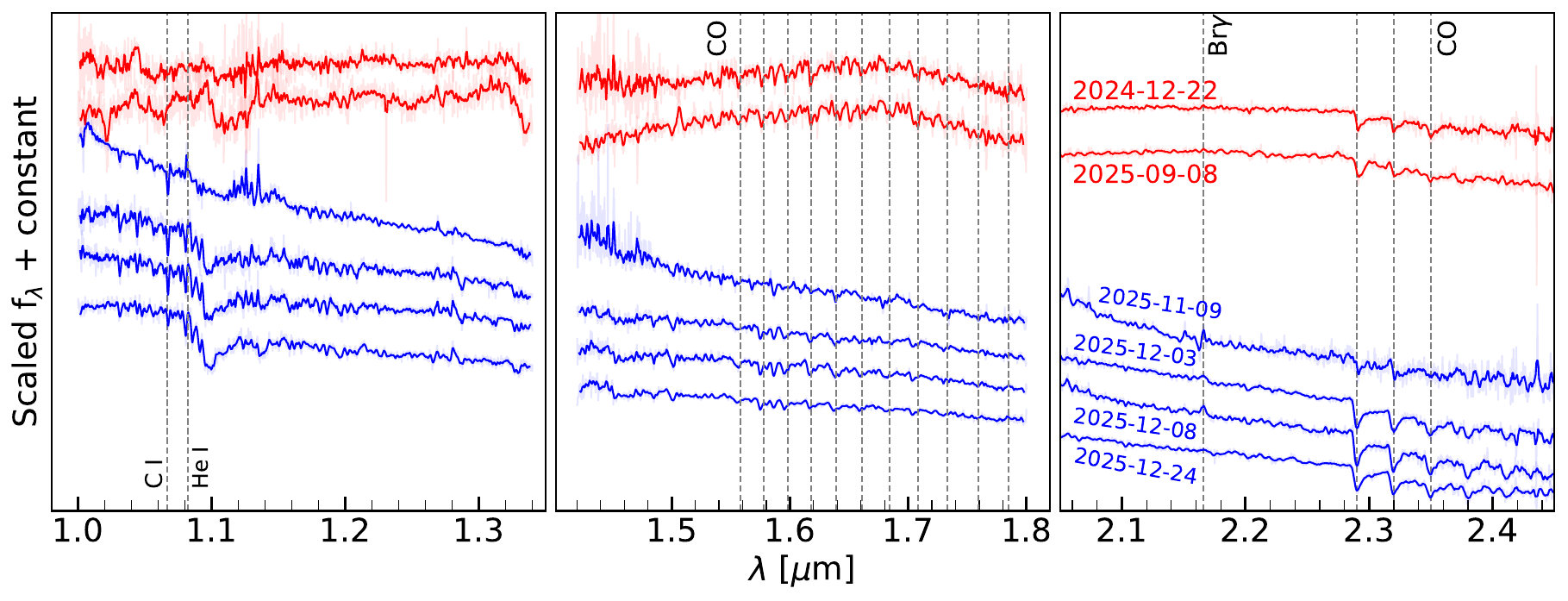}
    \caption{Near-infrared spectra of AT\,2025abao, from its precursor phase (red) and the eruption (blue). The spectra during the precursor phase resemble late M-type stars, while those during the eruption shows a transition to hotter temperatures. Notably, the CO absorption bands are seen in all spectra, indicating that some molecular gas remains intact even after the eruption.}
    \label{fig:all_nir_spectra}
\end{figure}

\begin{figure}[H]
    \centering
    \includegraphics[width=\linewidth]{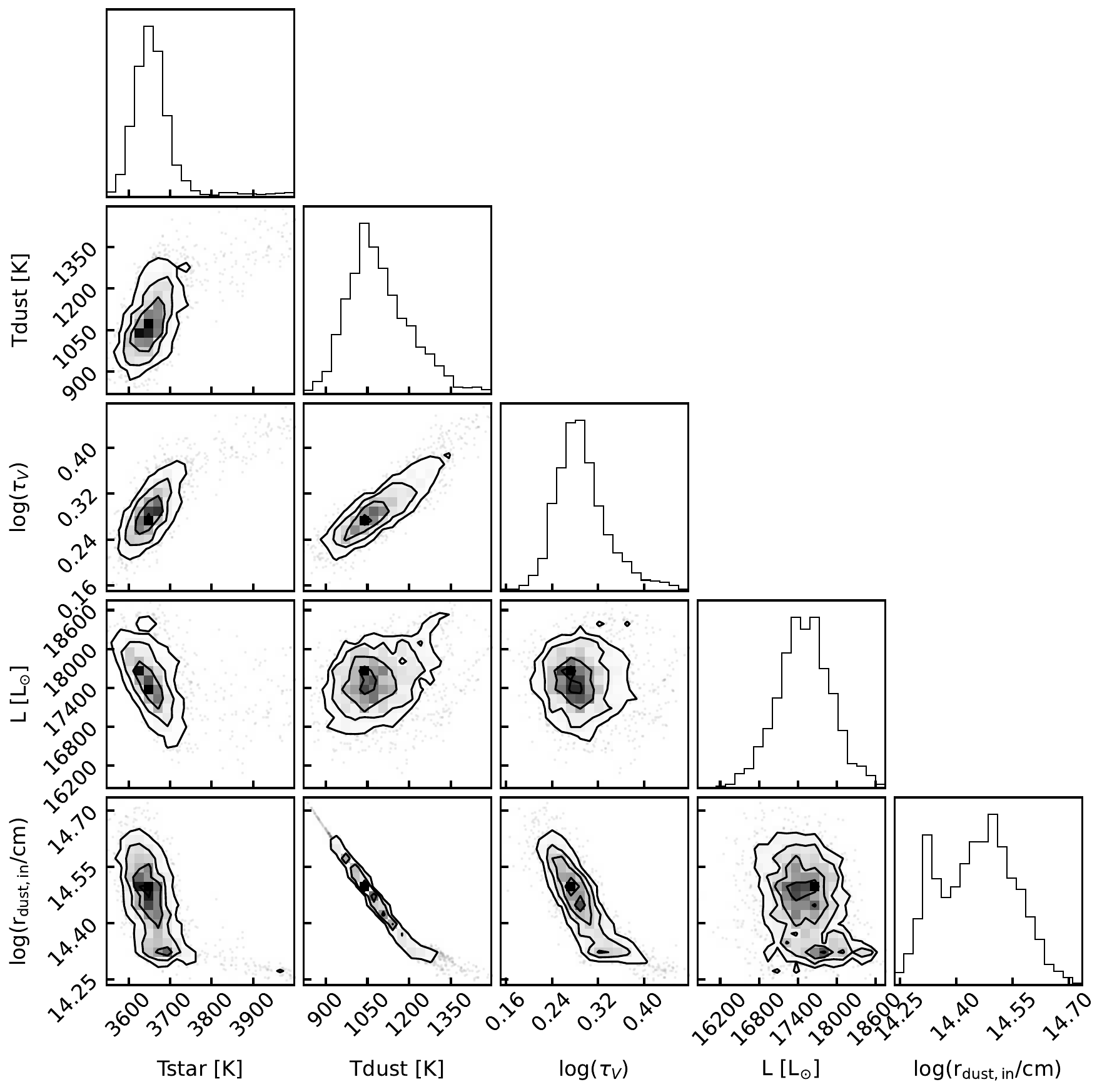}
    \caption{Posterior distributions of parameters inferred from SED modeling of the progenitor in Section\,\ref{sec:modeling_progenitor}.}
    \label{fig:corner_progenitor}
\end{figure}

\begin{figure}[H]
    \centering
    \includegraphics[width=\linewidth]{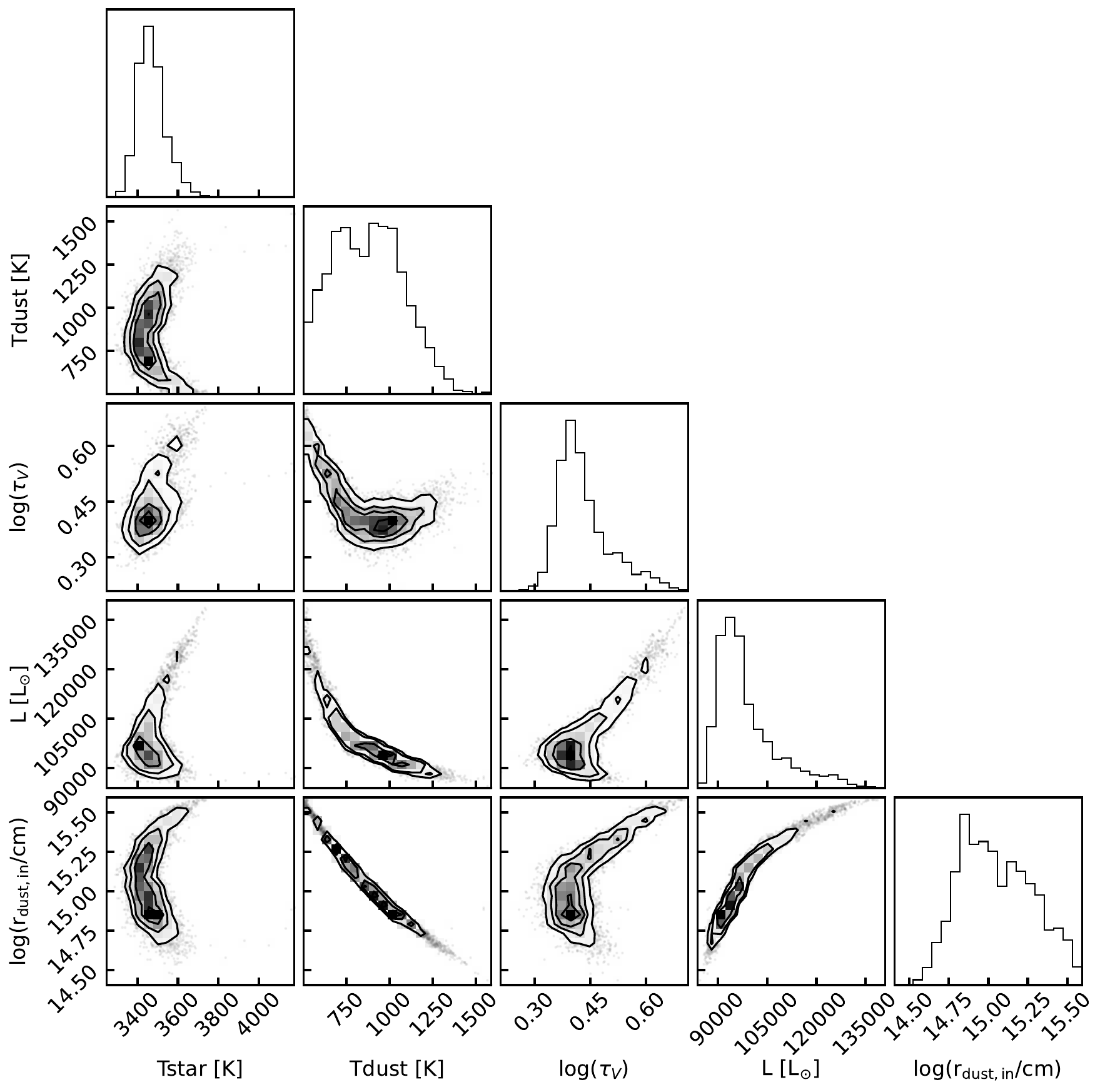}
    \caption{Posterior distributions of parameters inferred from SED modeling of the SPHEREx SED in Section\,\ref{sec:modeling_spherex}.}
    \label{fig:corner_spherex}
\end{figure}

\begin{figure}[H]
    \centering
    \includegraphics[width=\linewidth]{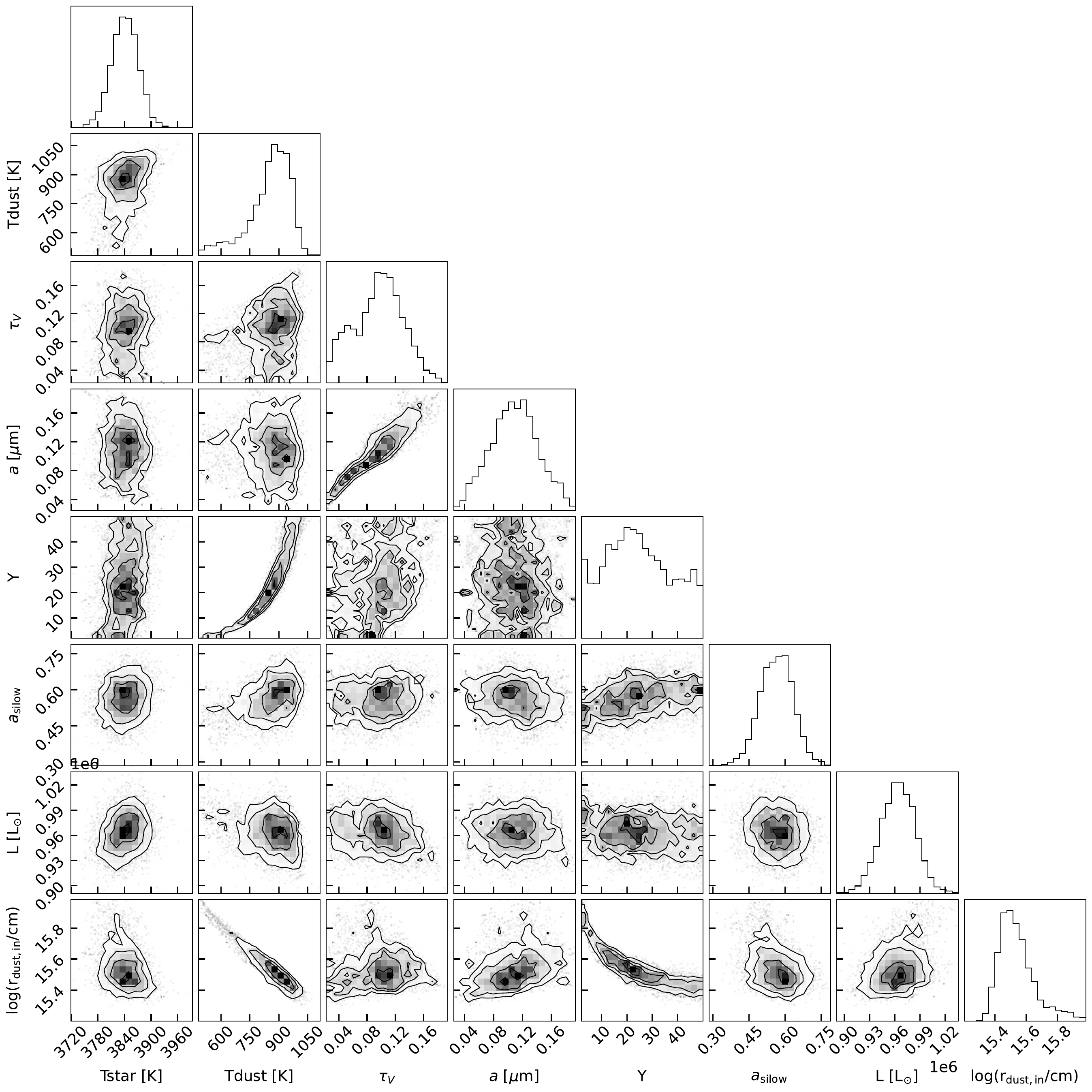}
    \caption{Posterior distributions of parameters inferred from SED modeling of the \emph{JWST} SED in Section\,\ref{sec:modeling_jwst}.}
    \label{fig:corner_jwst}
\end{figure}

\begin{figure}[H]
    \centering
    \includegraphics[width=\linewidth]{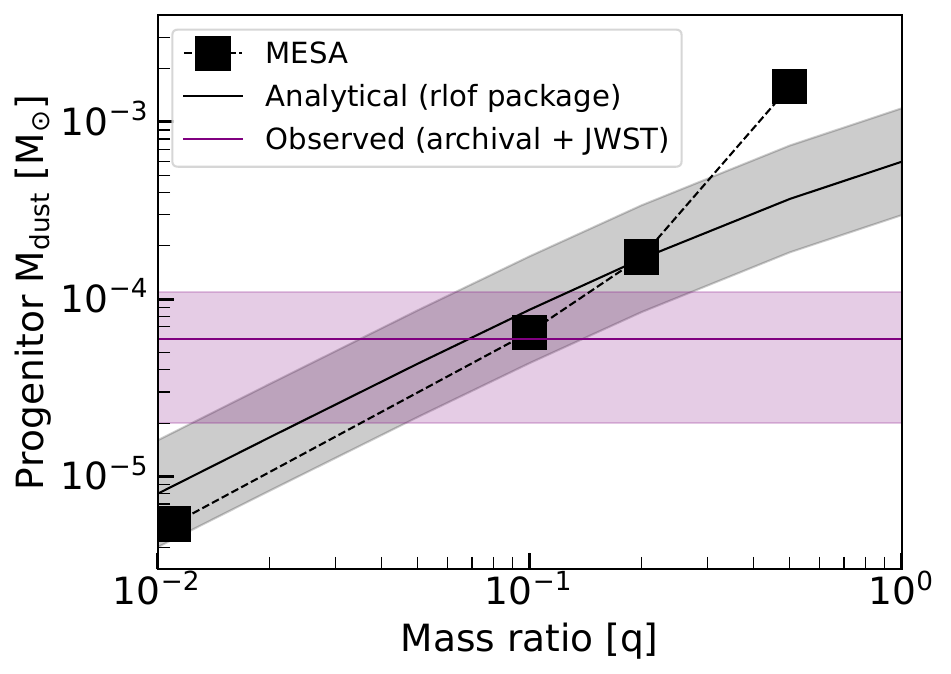}
    \caption{Theoretically calculated dust mass around the quiescent progenitor system as a function of binary mass ratio using analytical prescriptions (black line) and detailed simulations (squares). The black shaded region marks the range expected by varying stellar parameters in the analytical model (Section\,\ref{sec:disc_mass_loss_comparison}). The observed dust mass from the archival progenitor SED and the JWST observation is shown in purple, which constrains the system mass ratio to $q=0.10_{-0.08}^{+0.10}$.}
\label{fig:progenitor_mdust_comparisons}
\end{figure}

\begin{figure}[H]
    \centering
    \includegraphics[width=\linewidth]{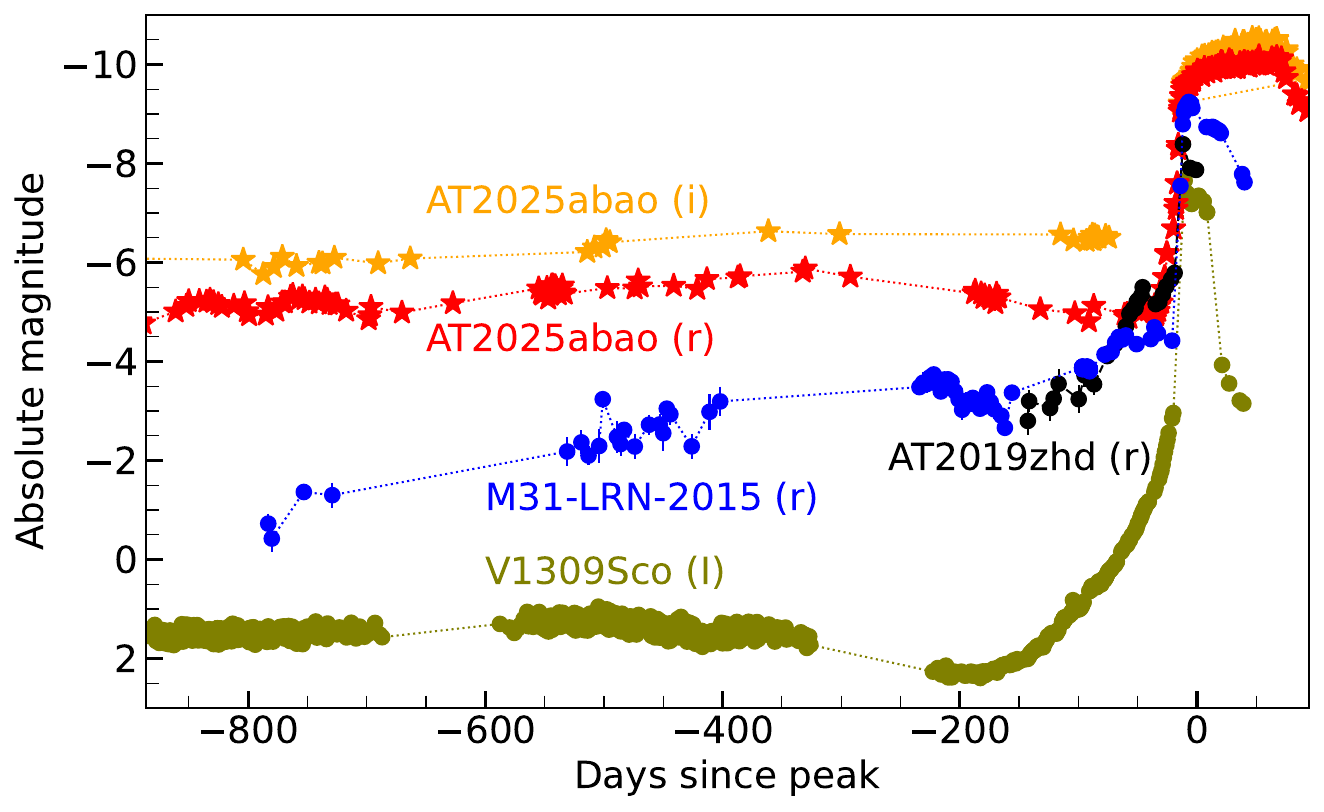}
    \caption{Optical lightcurves of LRNe with precursor detections. Previously identified sources such as V1309Sco, M31-LRN-2015, AT\,2019zhd (circles) exhibit a gradual brightening by several magnitudes prior to its eruption. AT\,2025abao (stars) differs from these sources, as it only shows a bumpy $\approx1$\,magnitude brightening before transitioning to its eruption.}
\label{fig:lc_precursor_comparisons}
\end{figure}

\begin{figure}[H]
    \centering
    \includegraphics[width=\linewidth]{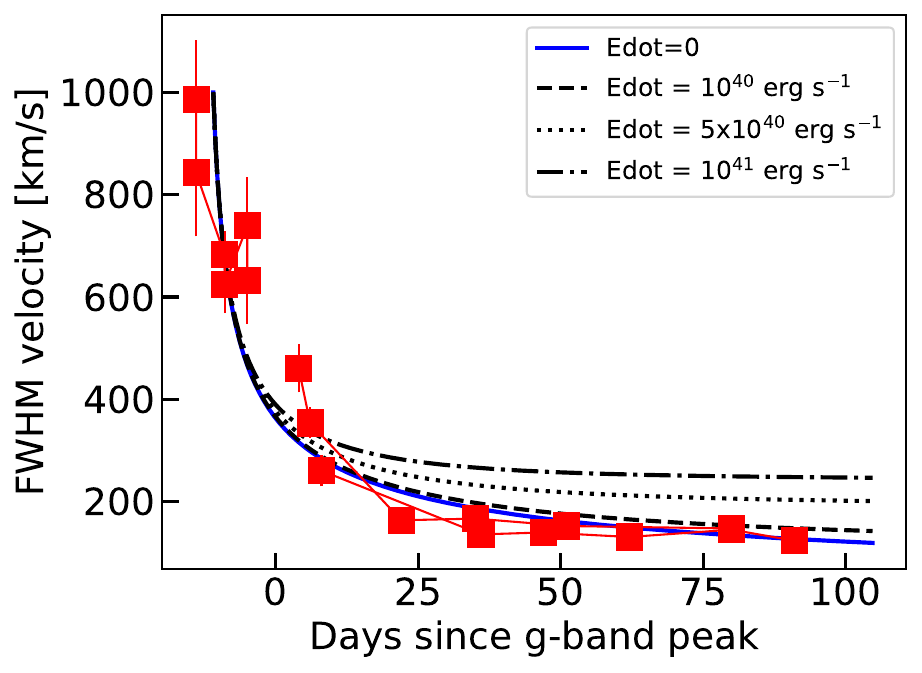}
    \caption{Evolution of FWHM velocities inferred from our spectra (red), compared to theoretical expectations for momentum-conserving collisions with CBM following a $\rho(r)\propto r^{-2}$ density profile (blue solid line). The broken lines indicate the expectation by including an additional mechanical power source, such as a surviving binary, injecting momentum into the ejecta. Our observations constrain the magnitude of this power source $\dot{E}\lesssim 10^{40}$\,erg\,s$^{-1}$.}
\label{fig:velocity_evolution_additional_energy}
\end{figure}

\begin{figure}[H]
    \centering
    \includegraphics[width=\linewidth]{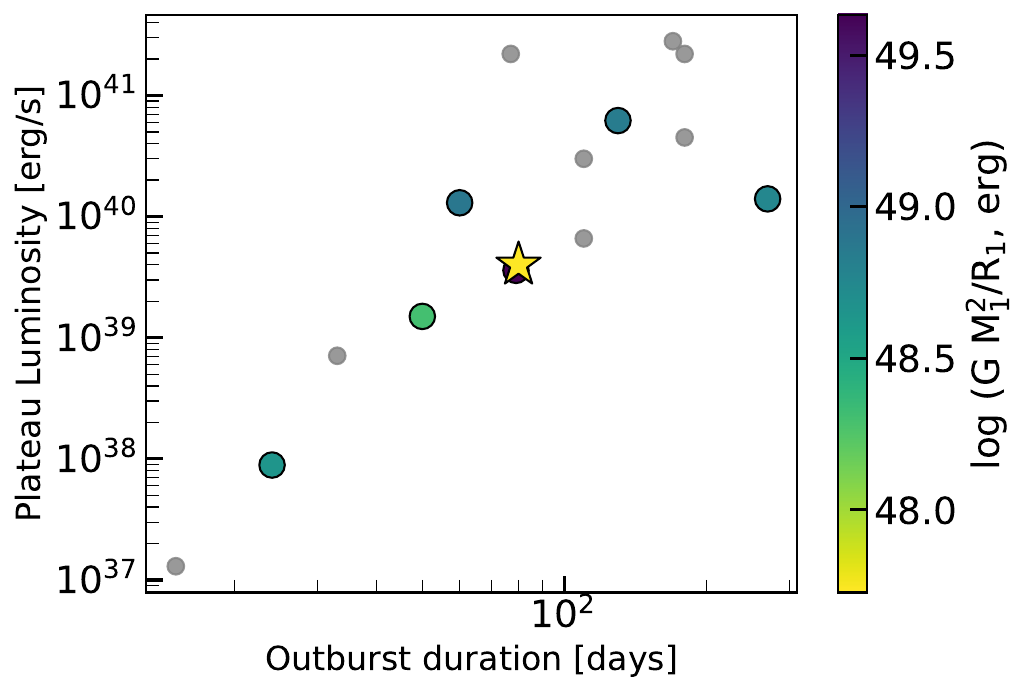}
    \caption{LRN plateau luminosity as a function of the plateau duration. LRNe for which progenitor primary stars have been identified are plotted as colored points, color-coded by the binding energy of the star. LRNe with no progenitor identifications are plotted as gray points. The eruption of AT\,2025abao (star) is remarkably similar to other LRNe, despite having a progenitor with the lowest binding energy ever measured.}
\label{fig:lrn_properties_comparison}
\end{figure}

\begin{figure}[H]
    \centering
    \includegraphics[width=\linewidth]{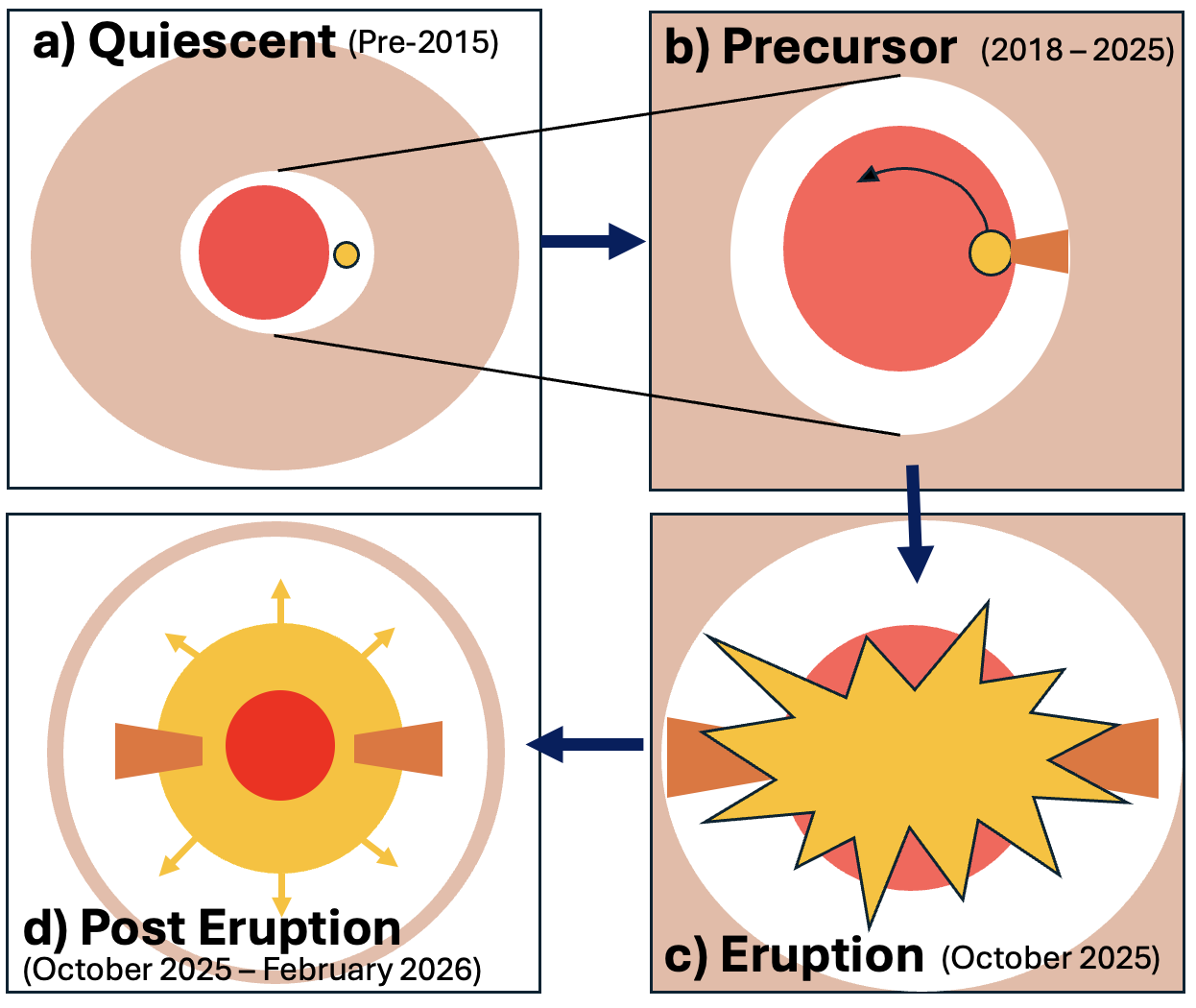}
    \caption{A graphical representation of the different phases of AT\,2025abao. a) The quiescent binary system comprising a $\approx7$\,M$_{\odot}$ AGB and a $\approx0.7$\,M$_{\odot}$ star, surrounded by an extended dust shell from mass loss from the binary following Roche Lobe overflow of the primary. b) The precursor brightening phase that began in 2018, marking the onset of CEE, powered by frictional drag on the inspiraling companion lasting for $\approx7$ years. The mass outflow rate during this phase is $\sim10^{-2}$\,M$_{\odot}$\,yr$^{-1}$, producing $\sim0.1$\,M$_{\odot}$ of CBM in an equatorial disk around the binary (dark brown), c) The ballistic eruption in late 2025 from the merger of the inspiraling companion with the primary core, resulting in the ejection of $\approx0.1-0.2$\,M$_{\odot}$ of the primary envelope, and sublimation of a large fraction of dust around the binary.  d) The eventual lightcurve powered by shock-interactions between freely expanding merger ejecta (indicated by yellow arrows) and equatorially focused CBM. The surviving pre-existing dust is traced by our \emph{JWST} observations. The black arrows between the panels indicate the progression of time.}
    
\label{fig:graphic}
\end{figure}

\end{document}